# Estimating the potential to prevent locally acquired HIV infections in a UNAIDS Fast-Track City, Amsterdam

Alexandra Blenkinsop†[1,2], Mélodie Monod[1], Ard van Sighem[3], Nikos Pantazis[4], Daniela Bezemer[3], Eline Op de Coul[5], Thijs van de Laar[6,7], Christophe Fraser[8], Maria Prins[9], Peter Reiss[2,10], Godelieve de Bree[2,11], Oliver Ratmann†[1] on behalf of the HIV Transmission Elimination Amsterdam project

[1] Department of Mathematics, Imperial College London, London
[2] Amsterdam Institute for Global Health and Development, Amsterdam, The Netherlands
[3] Stichting HIV Monitoring, Amsterdam, The Netherlands
[4] Department of Hygiene, Epidemiology and Medical Statistics, Medical School, National and Kapodistrian University of Athens
[5] Center for Infectious Diseases Prevention & Control, RIVM
[6] Department of Donor Medicine Research, laboratory of Blood-borne Infections, Sanquin Research
[7] Department of Medical Microbiology, Onze Lieve Vrouwe Gasthuis, Amsterdam, The Netherlands
[8] Big Data Institute, Li Ka Shing Centre for Health Information and Discovery, Nuffield Department of Medicine, University of Oxford, Oxford
[9] Academic Medical Center - Academisch Medisch Centrum, Amsterdam
[10] Department of Global Health, Amsterdam University Medical Centers, University of Amsterdam
[11] Division of Infectious Diseases, Department of Internal Medicine, Amsterdam Infection and Immunity Institute

† Corresponding author. Email: a.blenkinsop@imperial.ac.uk, oliver.ratmann@imperial.ac.uk

## Abstract

Amsterdam and other UNAIDS Fast-Track cities aim for zero new HIV infections. Utilising molecular and clinical data of the ATHENA observational HIV cohort, our primary aims are to estimate the proportion of undiagnosed HIV infections and the proportion of locally acquired infections in Amsterdam in 2014-2018, both in MSM and heterosexual individuals and Dutch-born and foreign-born individuals.

We located diagnosed HIV infections in Amsterdam using postcode data (PC4) at time of registration in the ATHENA observational HIV cohort, and estimated their date of infection using clinical HIV data. We then inferred the proportion of undiagnosed from the estimated times to diagnosis. To determine the sources of Amsterdam infections, we used HIV sequences of Amsterdam people living with HIV (PLHIV) within a background of other Dutch and international sequences to phylogenetically reconstruct transmission chains, and tabulate their growth between 2014 and 2018. Frequent late diagnoses indicate that more recent phylogenetically observed chains are increasingly incomplete, and we use a Bayesian model to estimate the actual growth of Amsterdam transmission chains, and the proportion of locally acquired infections.

We estimate that 20% [95% CrI 18-22%] of infections acquired among MSM in Amsterdam between 2014-2018 were undiagnosed by the start of 2019, and 44% [37-50%] among heterosexuals, with variation by place of birth. In this period, the estimated proportion of Amsterdam MSM infections in 2014-2018 that were locally acquired was 68% [61-74%], with no substantial differences by region of birth. In Amsterdam heterosexuals, this was 57% [41-71%] overall, with heterogeneity by place of birth.

The data indicate substantial potential to further curb local transmission, in both MSM and heterosexual Amsterdam residents. In 2014-2018 the largest proportion of local transmissions in Amsterdam are estimated to have occurred in foreign-born MSM, suggesting foreign-born MSM would likely benefit most from intensified interventions.

# Introduction

Human immunodeficiency virus (HIV) is concentrated in metropolitan areas, with the 200 cities with the highest burden of HIV representing 26% of global HIV burden (Joint United Nations Programme on HIV/AIDS (UNAIDS) 2014). In response, as of March 2021 over 300 cities have joined the Fast-Track Cities initiative by signing the Paris Declaration, committing to end the AIDS epidemic by 2030, by addressing disparities in access to basic health and social services, social justice and economic opportunities (UNAIDS 2019). Several of these fast-track cities have successfully developed strategies which best address the needs of the local epidemic, including London's HIV Prevention Programme and early ART initiation, and New York's Status Neutral Prevention and Treatment Cycle (Public Health England 2018; Myers et al. 2018). A central milestone in this agenda is to characterise the number of HIV infections that are acquired from sources within cities and are thus preventable through local interventions, as well as to identify the primary risk groups with infections from local sources.

In the Netherlands, Amsterdam is the city with the greatest HIV burden nationally, reflecting in part a large MSM community, as well as large communities of at-risk, foreign-born individuals. Amsterdam has a long history of a collaborative HIV approach in combating the epidemic and joined the UNAIDS fast-track Cities initiative on 1 December 2014. City-level HIV responses were galvanised in the HIV Transmission Elimination Amsterdam project (H-Team) that same year (de Bree et al. 2019). The H-Team fast-track response, amongst others, focussed on outreach activities, encouraging repeat testing every 3-6 months to identify acute and early HIV infection, followed by immediate initiation of combination antiretroviral therapy (c-ART) in newly diagnosed patients, and roll-out of pre-exposure prophylaxis (PreP) in high risk populations at increased risk of HIV infection (den Daas et al. 2018; Bartelsman et al. 2017; Hoornenborg et al. 2019; M. Dijkstra et al. 2019). Prior to the COVID-19 pandemic, the number of annual HIV diagnoses in Amsterdam residents has consistently declined from 300 new city-level HIV diagnoses in 2010 to 120 in 2018, primarily in Dutch-born and foreign-born MSM. Given these impressive achievements, it is particularly unclear how many new infections are locally acquired and could thus still be locally averted. Late diagnoses remain common and are a particular concern in this effort, both for individual health and the risk that unnoticed transmission chains pose to public health.

Here, we build on Amsterdam's combined case and genomic surveillance data to reconstruct transmission chains at city level, and the growth and origins of these chains between 2014-2018. We estimate the extent of undiagnosed infections at the forefront of the cities' transmission chains, among infections that are estimated to have occured since 2014. We characterise variation in the extent to which all individuals in city level transmission chains are virally suppressed, and we study the relative impact of newly introduced transmission chains from outside Amsterdam on the city-level epidemic. Finally, we combine our insights to estimate the proportion of locally acquired infections, i.e. those infections that could have been locally averted.

## Methods

Demographic and clinical cohort data comprising city-level infections

Data were obtained from the prospective ATHENA cohort of all people living with HIV (PLHIV) in care in the Netherlands, including patient demographics and longitudinal CD4, HIV viral load, viral sequence, and treatment data (see Supplementary Material, Section 2) (Boender et al. 2018). Sequencing methods are described in (Bezemer et al. 2004). We geolocated diagnosed infections in Amsterdam based on patients' postcode of residence at time of first registration in ATHENA, or the most recent registration update. MSM were stratified by region of birth (The Netherlands; Western Europe, North America, Oceania; Eastern and Central Europe; South America and the Caribbean; Other), and similarly for heterosexual individuals (The Netherlands; South America and the Caribbean; Sub-Saharan Africa; Other), and so we considered 9 risk groups in total.

We here focus on city-level transmission chains growing in the period from January 1, 2014 to December 31, 2018, which for brevity we refer to as 2014-2018. Available demographic, clinical, and viral sequence data were obtained for HIV diagnoses in Amsterdam until December 31, 2018, from the ATHENA database version closed on May 1, 2019.

Estimating HIV infection dates, and undiagnosed infections

Using longitudinal viral load and CD4 count data along with other demographic/clinical information in combination with parameters of a bivariate linear mixed model, estimated in a large dataset of individuals with known infection dates from the CASCADE Collaboration, we estimated time from infection to diagnosis for all HIV diagnosed patients using a Bayesian method. (Pantazis et al. 2019). We next reconstructed time-to-diagnosis distributions from the individual-level estimates. To avoid censoring of infection-to-diagnosis times, we focused analyses on the subset of infections in 2010-2012 which were diagnosed by May 2019, since most infections in this window would have been diagnosed by the close of study. We assume time to diagnosis did not change substantially in the years 2010-2019. We then fitted a Bayesian hierarchical model with a Weibull likelihood, borrowing information across individuals stratified by region of birth. The model was implemented with Stan version 2.21 (Carpenter et al. 2017). Full details are provided in Supplementary Material, Section 3.

For each year since 2014, we calculated the probability that infections were not diagnosed by database closure, and used the average of these probabilities to estimate the proportion of infections in 2014-2018 that remained undiagnosed by database closure, which we denote by θ. The total number of Amsterdam infections in 2014-2018 including the undiagnosed, which we denote by $N_I$, was calculated by dividing the number of diagnosed Amsterdam infections in 2014-2018, $N_D$, with the estimated proportion of diagnosed individuals through,

$$N_I = \frac{N_D}{1-\theta}. \qquad (1)$$

We calculate the proportion of infections in 2014-2018 that remain undiagnosed and (1) for Amsterdam MSM and heterosexuals, and region of birth.

Phylogenetic reconstruction of city-level transmission chains

To reconstruct distinct transmission chains among city-level infections, we used the first available partial HIV-1 *polymerase* (*pol*) sequence from Amsterdam PLHIV, Dutch PLHIV from outside Amsterdam, and >82,000 *pol* sequences from non-Dutch PLHIV that were at least 1300 nucleotides long from other countries classified into 10 regions: Africa, Western Europe, Eastern Europe and Central Asia, North America, Latin America and the Caribbean, Dutch Caribbean and Suriname, Middle East and North Africa, Asia and Oceania. The non-Dutch viral sequences were retrieved from the Los Alamos HIV-1 sequence database on March 2, 2020. All sequences were subtyped using Comet (Struck et al. 2014) and Rega (Pineda-Peña et al. 2013). Subtype-specific alignments were generated with *Virulign (Libin et al. 2019)* (Supplementary Text Section 4.1).

Subtype-specific HIV phylogenetic trees were reconstructed with FastTree v2.1.8 (Price, Dehal, and Arkin 2010). Then, we attributed to all viral lineages in the phylogenies a 'state' label that included information on the transmission risk group (MSM, heterosexual, other) and place of birth (defined above) with *phyloscanner* version 1.8.0 (Wymant et al. 2018) and as in (Bezemer et al. 2022). In this analysis, lineages are grouped into phylogenetic subgraphs that have the same, uninterrupted state label based on maximum parsimony. Diagnosed Amsterdam patients in the same subgraph are interpreted as belonging to the same transmission chain, and the estimated state of the root of the subgraph is interpreted as the origin of the transmission chain. We here refer to the subgraphs also as the phylogenetically observed (parts of) transmission chains. To capture phylogenetic uncertainty, phylogenetic analyses were repeated on 100 bootstrap replicates drawn from each subtype alignment, and transmission chains were enumerated across these replicate analyses. See Supplementary Text 4.2 for full details.

We classified phylogenetically reconstructed transmission chains by the infection dates that we estimated from each patient's diagnosis date, risk group, age, CD4 trajectory and viral load trajectory. Chains were classified as 'pre-2014' if at least one of its members had a posterior median infection date before 2014, and as 'emerging' if all members had a posterior median infection date after January 1, 2014.

Virally unsuppressed transmission chains

For all pre-2014 chains, we determine the number of infectious individuals at the start of 2014 from viral load data. Specifically, we defined patients as suppressed by 2014 if their last viral load measurement before 2014 was below 100 copies/ml, and count for each pre-2014 chain its suppressed and unsuppressed members by 2014.

Estimating the growth of city-level transmission chains

Because of the large number of late presenters and incomplete sequence coverage in diagnosed patients, the phylogenetically observed transmission chains are incomplete and statistical models were required to estimate the growth and origins of Amsterdam transmission chains. We extended the Bayesian branching process model of (Bezemer et al. 2022) to describe the growth of pre-existing transmission chains and transmission chains introduced since 2014. In the extension, the model likelihood is described by the observed number of new cases in each chain conditional on the number of index cases, and integrates out all possible scenarios of unobserved members of the actual transmission chains (Supplementary Text Section 5). In the model, the index cases are assumed to be infectious and defined by the number of unsuppressed members by 2014, adjusted for the sampling probability of such members. In chains which emerged since 2014, we assume that there is one index case. The likelihood then comprises the growth distributions of emerging chains (since 2014 as defined above), pre-2014 chains that continued to grow, and pre-2014 chains with unsuppressed members that did not grow. Pre-2014 chains for which all members were suppressed by 2014 and which did not grow were not included, because these chains had no unsuppressed index case. We fitted the branching process model with Stan version 2.21 to MSM chains borrowing information across subtypes, and similarly for heterosexual individuals. The primary output of the model are posterior predictive distributions on the growth of the actual transmission chains while accounting for as of yet undiagnosed and unsequenced individuals. This includes emerging chains that were entirely unsampled. Full details are provided in the Supplementary Text, Section 6.

Derived statistical estimates

Estimates of the proportion of Amsterdam infections between 2014-2018 that originated from an individual living in Amsterdam, can be calculated from the predicted actual transmission chains. We denote this proportion by $\gamma$, and interpret it as the proportion of infections which have the potential to be locally preventable through local intervention. This is because all infections originating from an individual living in Amsterdam had a local source, except the index cases in the emerging chains. We have,

$$\gamma = 1 - \frac{N_C}{N_I}, (2)$$

where $N_C$ is the estimated number of transmission chains which emerged between 2014-2018, and $N_I$ is the estimated number of new infections between 2014-2018 in both pre-2014 chains and emerging transmission chains including the index case. Several emerging chains in heterosexuals had a phylogenetically likely origin in Amsterdam MSM, prompting us to adjust Equation 2 accordingly.

Through Equation 2, we obtain estimates of $\gamma$ for both MSM and heterosexuals, and for each subtype. To obtain estimates of transmission chains and of $\gamma$ that are stratified by place of birth, we used weighted averages across chains and subtypes, with the weight determined as the proportion of observed individuals from a particular area of birth. Full details are provided in the Supplementary Text, Section 6. Figure 1 illustrates how each of the data sources are used to build the model and estimate the proportion, and number, of locally acquired infections.

Ethics

As from 2002 ATHENA is managed by SHM, the institution appointed by the Dutch Ministry of Public health, Welfare and Sport for the monitoring of people living with HIV in the Netherlands. People entering HIV care receive written material about participation in the ATHENA cohort and are informed by their treating physician on the purpose of data collection, thereafter they can consent verbally or elect to opt-out. Data are pseudonymised before being provided to investigators and may be used for scientific purposes. A designated data protection officer safeguards compliance with the European General Data Protection Regulation. (Boender et al. 2018).

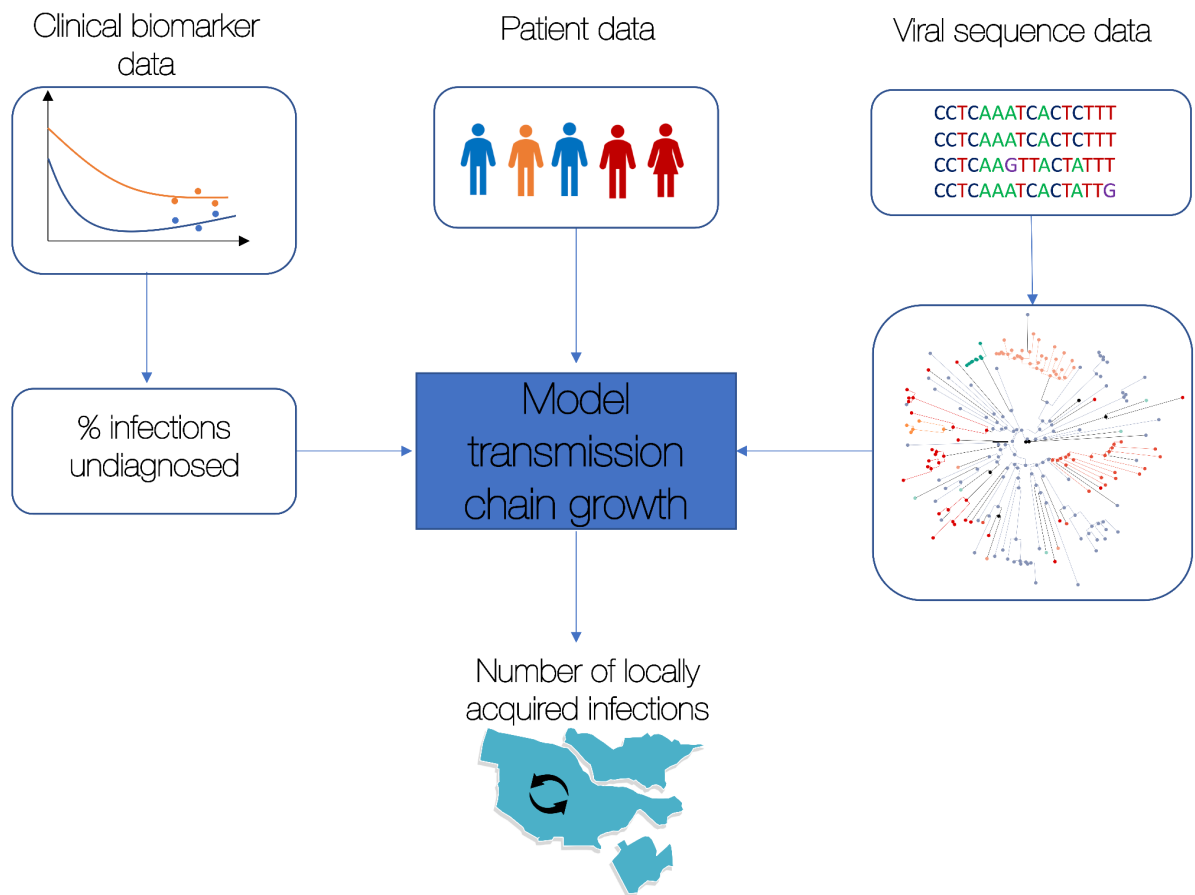

*Figure 1: Graphic describing approach to analysis. Input data includes patient baseline data at registration, clinical biomarker data and viral sequence data. Biomarker data is used to estimate the proportion of undiagnosed infections, and thus the total population size of PLHIV. Sequence data is used to reconstruct phylogenetic trees. The phylogenetically likely subgraphs are extracted and used to model the growth of transmission chains in Amsterdam, adjusting for the sequence sampling fraction.*

## Results

Substantial declines in HIV diagnoses and infections in Amsterdam

Between January 1 2014 and May 1 2019, there were 846 HIV diagnoses in Amsterdam residents who self-identified as MSM (79%) or heterosexual (21%). 275 (33%) of these diagnoses presented with a CD4 count below 350, with late presentation being higher among heterosexuals. All diagnosed patients had biomarker data available to estimate time to diagnosis, and 516 of 846 (61%) were estimated to have been infected between 2014-2018 based on the posterior median infection time estimate (Table 1). In the preceding five-year period 2009-2013, there were 1436 HIV diagnoses in Amsterdam and 1128 diagnoses with estimated infection in 2009-2013, suggesting a substantial reduction in infections in 2014-2018. Yet, the rate of new Amsterdam diagnoses since 2014 (104 per 100,000) remained higher than the national rates excluding Amsterdam (24 per 100,000), and in this sense Amsterdam remains a HIV hotspot in the Netherlands.

| Risk group | Amsterdam diagnosed 2014-May 2019 | Diagnosed with CD4<350 | Estimated infected 2014-2018 | Estimated time to diagnosis† | Estimated undiagnosed until May 2019 | Estimated total infected 2014-2018 |
|---|---|---|---|---|---|---|
| | (n) | (n) | (n) | (years) | (%) | (n) |
| MSM (Dutch-born) | 298 | 103 | 190 | 0.71 [0.61-0.81] | 17% [15-20%] | 229 [223-237] |
| MSM (Born in W. Europe, N. America and Oceania) | 100 | 12 | 80 | 0.67 [0.51-0.86] | 16% [11-21%] | 95 [90-101] |
| MSM (Born in E. and C. Europe) | 51 | 8 | 32 | 0.91 [0.66-1.32] | 22% [16-32%] | 41 [38-47] |
| MSM (Born in S. America and the Caribbean) | 124 | 38 | 83 | 0.93 [0.72-1.2] | 23% [19-30%] | 108 [102-118] |
| MSM (Born in any other country) | 98 | 31 | 61 | 1.12 [0.80-1.53] | 27% [20-34%] | 83 [76-93] |
| **MSM (all)** | **671** | **192** | **446** | **0.83 [0.61-1.14]** | **20% [18-22%]** | **557 [543-575]** |
| Heterosexuals (Dutch-born) | 51 | 19 | 23 | 1.49 [0.94-2.19] | 34% [23-47%] | 35 [30-43] |
| Heterosexuals (Born in Sub-Saharan Africa) | 67 | 36 | 17 | 3.25 [2.49-3.97] | 60% [48-69%] | 42 [33-55] |
| Heterosexuals (Born in S. America and the Caribbean) | 37 | 18 | 21 | 1.43 [0.92-2.12] | 34% [23-47%] | 35 [30-43] |
| Heterosexuals (Born in any other country) | 20 | 10 | 9 | 2.19 [1.39-3.23] | 44% [31-59%] | 16 [13-22] |
| **HSX (all)** | **175** | **83** | **70** | **2.02 [1.23-3.22]** | **44% [37-50%]** | **124 [111-141]** |
| **Total** | **846** | **275** | **516** | - | **24% [22-27%]** | **682 [662-705]** |

† *Posterior estimated median time from infection to diagnosis [95% CI]*

*Table 1: Summary of individuals diagnosed between 2014 to May 2019, late presenters, with an estimated infection date between 2014-2018, and total estimated infected.*

Most Amsterdam diagnoses and infections are in foreign-born MSM

190 (37%) Amsterdam diagnoses with estimated infection in 2014-2018 were in Dutch-born MSM, 256 (50%) in foreign-born MSM, 23 (4%) in Dutch-born men and women identifying as heterosexuals, and 54 (9%) in foreign-born heterosexuals. Thus, the largest proportion of new diagnoses in Amsterdam with infection dates between 2014-2018 were in foreign-born MSM. 2% of individuals opt out of the ATHENA study (Boender et al. 2018), and 5.2% of individuals were lost to care (Boender et al. 2018; van Sighem et al. 2020).

Overall, we find the individual-level time-to-diagnosis estimates varied substantially within each of the 9 risk groups by transmission group and region of birth (Supplementary Figures 1 and 2). Regardless, the posterior median time-to-diagnosis estimates among individuals were 14 months longer in heterosexuals than in MSM, and 9 months longer among heterosexuals born in the Netherlands than Dutch-born MSM. Among heterosexuals, median times to diagnosis were 21 months longer among heterosexuals born in Sub-Saharan Africa compared to Dutch-born heterosexuals (Table 1). While the estimated times to diagnosis are shortening compared to earlier calendar periods, the substantial diagnosis delays continue to undermine the long-term prognosis of infected individuals and transmission prevention efforts.

High proportion of infections since 2014 that remained undiagnosed by May 2019

Local estimates of the continuum of care indicate that Amsterdam has surpassed the 95-95-95 targets, with an estimated 5% of all people in Amsterdam living with HIV that remained undiagnosed by the end of 2019 (van Sighem et al. 2020; UNAIDS 2019). Based on the time-to-diagnosis estimates in our cohort, we can quantify the proportion of recent Amsterdam infections in 2014-2018 that remained undiagnosed by May 1 2019. Figure 2 shows that the estimated undiagnosed proportions are considerably higher when we focus on infections acquired since 2014. On average, over the entire study period, an estimated 20% [18-22%] of infections in MSM in 2014-2018 remained undiagnosed by December 2018, and 44% [37-50%] in heterosexuals (Table 1). The highest proportion of undiagnosed Amsterdam infections since 2014 are estimated in heterosexuals born in Sub-Saharan Africa, with 60% [48-69%].

The bivariate model of biomarker data that underpins these estimates has been validated in simulation studies (Pantazis et al. 2019), however we further explored in sensitivity analyses the implications of upwards bias in the time-to-diagnosis estimates. Repeating the calculations based on the 40% and 30% lower quantiles of time-to-diagnoses estimates, we find that 21% [19-23%] and 16% [15-18%], respectively, of Amsterdam infections in 2014-2018 are estimated to be undiagnosed by May 2019. Further details are presented in the Supplementary Material, Section 3.3.

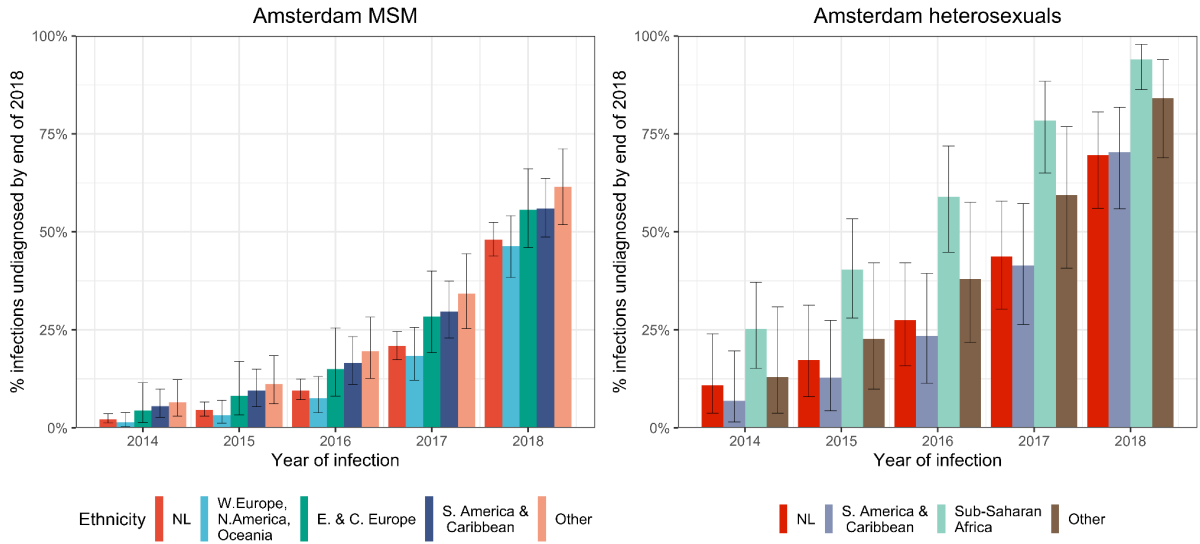

*Figure 2: Estimated percentage of infections between 2014-2018 which remained undiagnosed by May 2019, by year of infection with 95% credible intervals.*

Substantial variation in city-level transmission chains that are virally suppressed

We next adopted viral phylogenetic methods to understand how the diagnosed Amsterdam infections since 2014 are distributed across Amsterdam's HIV transmission networks. 387 of the 516 (75%) individuals had a partial *polymerase* HIV sequence available, of whom 344 were of the major subtypes or circulating recombinant forms that are circulating in Amsterdam (B, 01AE, 02AG, C, D, G, A1 or 06cpx). 43 individuals were excluded from further analysis as they were associated with other subtypes or recombinant forms, or their subtype identification was inconclusive. Supplementary Table S1 summarises the characteristics of the study population, and those with a sequence available. We reconstructed viral phylogenies using the HIV sequence data from these individuals combined with viral sequences from 3,647 Amsterdam diagnoses with estimated infection prior to 2014, 6,875 diagnosed individuals from the Netherlands outside Amsterdam, and 14,222 viral sequences from outside the Netherlands that were genetically closest to those circulating in the Netherlands.

We identified across the major HIV-1 subtypes and circulating recombinant forms 1,829 distinct viral phylogenetic subgraphs that comprised at least one diagnosed Amsterdam infection prior to 2014, which we refer to as the phylogenetically observed pre-2014 chains (Figure 3 and Supplementary Figure S16). There were 1,253 pre-2014 chains in MSM, of which 949 (76%) had all members virally suppressed as of 2014, and of those 906 (95%) had no new member in 2014-2018. The remaining 5% of subgraphs likely grew from unsuppressed index individuals that were not sampled. In heterosexuals, there were 576 pre-2014 chains, of which 401 (70%) had all members virally suppressed as of 2014, and of those 391 (98%) had no new member in 2014-2018. Thus, transmission appears to have stopped since 2014 in almost all phylogenetically observed pre-2014 chains that had all their observed members suppressed by 2014. However, the extent to which the phylogenetically observed pre-2014 city-level chains were virally suppressed differed substantially across risk groups (Supplementary Table 2).

Growth of the phylogenetically observed parts of city-level transmission chains

Considering growth, 89 (7%) of the 1,253 phylogenetically observed pre-2014 chains in Amsterdam MSM had at least one new member diagnosed in 2014-2018, and 114 chains emerged (Figure 2 and Table 2). In Amsterdam heterosexuals, 15 (3%) of the 576 phylogenetically observed pre-2014 chains had at least one new member diagnosed in 2014-2018, and 26 chains emerged. The emerging chains thus outnumbered the growing pre-2014 chains in both Amsterdam MSM and heterosexuals. However, in MSM there were 158 new observed members in the growing pre-2014 chains and 139 new observed members in emerging chains, while in heterosexuals there were 17 new observed members in the growing pre-2014 chains and 27 new observed members in emerging chains (Table 3).

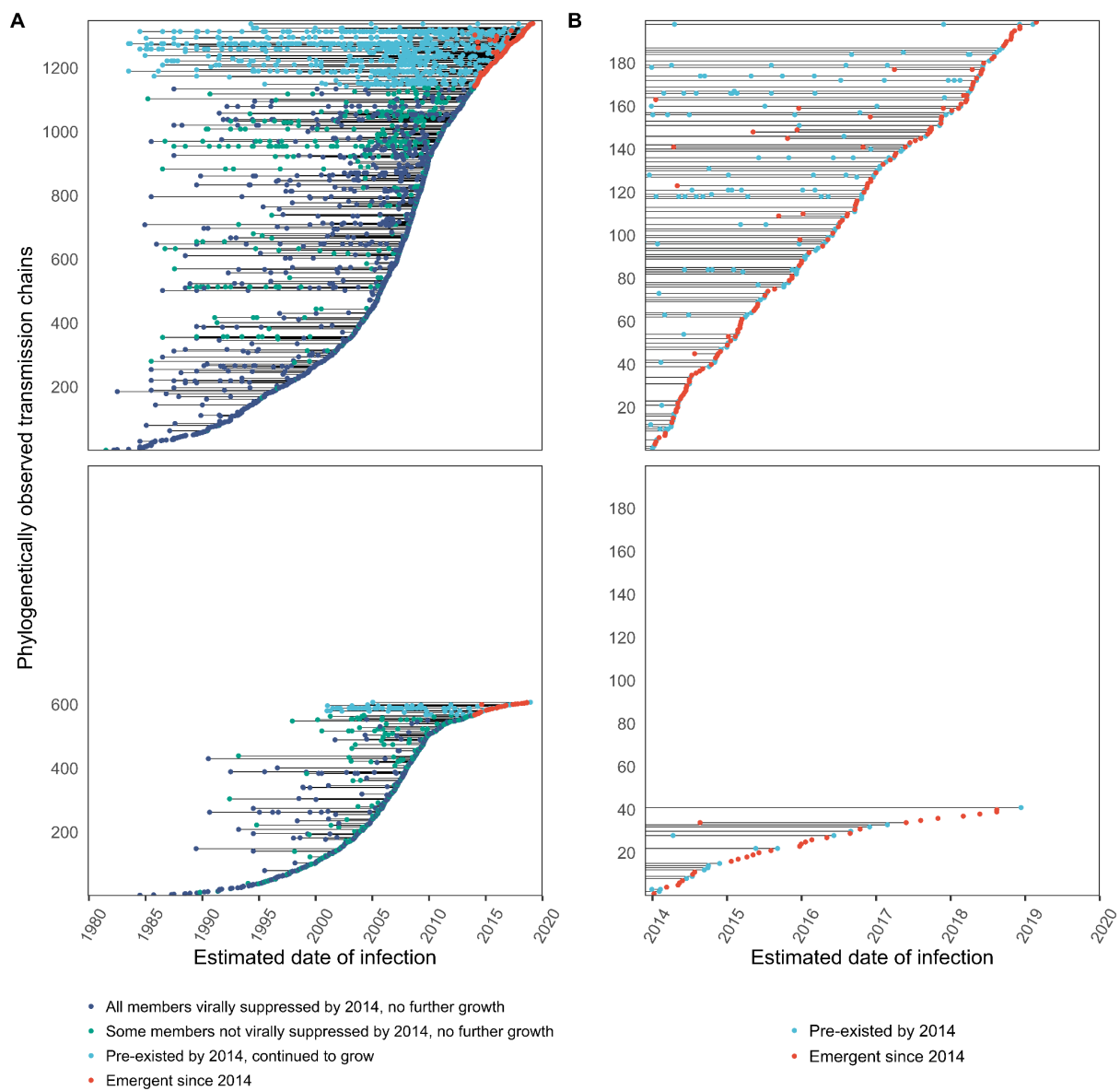

Figure 3: Phylogenetically observed parts of Amsterdam transmission chains. A) All chains. Horizontal lines connect individuals in reconstructed transmission chains in Amsterdam by chains which had no *new case since 2014, and those which continued to grow or emerged, among MSM (top) and heterosexuals (bottom), in order of last diagnosis per chain. B) Subset of chains with at least one individual estimated to have been infected since 2014. Data are presented as in subfigure A.*

| Transmission group | New cases | Pre-2014 chains Observed† (N) | (%) | Predicted‡ (N) | (%) | Emerging chains Observed† (N) | (%) | Predicted‡ (N) | (%) |
|---|---|---|---|---|---|---|---|---|---|
| Amsterdam MSM | 0 | 220 | 71.2% | 195 [170-218] | 63.1% [55.0-70.6%] | - | - | - | - |
| | 1 | 59 | 19.1% | 52 [37-69] | 16.8% [12.0-22.3%] | 94 | 82.5% | 144 [122-169] | 79.2% [71.5-85.9%] |
| | 2 | 15 | 4.9% | 24 [15-35] | 7.8% [4.9-11.3%] | 11 | 9.6% | 21 [12-31] | 11.4% [6.6-17.1%] |
| | 3 | 6 | 1.9% | 13 [6-21] | 4.2% [1.9- 6.8%] | 7 | 6.1% | 8 [3-14] | 4.2% [1.6- 7.6%] |
| | 4 | 3 | 1.0% | 8 [3-14] | 2.6% [1.0- 4.5%] | 2 | 1.8% | 4 [1-8] | 2.0% [0.5- 4.5%] |
| | 5 | 2 | 0.6% | 5 [1-10] | 1.6% [0.3- 3.2%] | 0 | 0.0% | 2 [0-6] | 1.1% [0.0- 3.0%] |
| | 6 | 0 | 0.0% | 3 [0-7] | 1.0% [0.0- 2.3%] | 0 | 0.0% | 1 [0-4] | 0.6% [0.0- 2.2%] |
| | 7+ | 4 | 1.3% | 8 [2-15] | 2.6% [0.6- 4.9%] | 0 | 0.0% | 2 [0-6] | 1.1% [0.0- 3.5%] |
| | Total that grew | 89 | | 114 [91-139] | | 114 | | 182 [161-207] | |
| | Total | 309 | | 309 [309-309] | | 114 | | 182 [161-207] | |
| Amsterdam heterosexual | 0 | 150 | 90.9% | 137 [122-150] | 83.0% [73.9-90.9%] | - | - | - | - |
| | 1 | 13 | 7.9% | 18 [9-29] | 10.9% [5.5-17.6%] | 25 | 96.2% | 52 [36-75] | 85.7% [74.1-94.9%] |
| | 2 | 2 | 1.2% | 5 [1-11] | 3.0% [0.6- 6.7%] | 1 | 3.8% | 6 [2-12] | 9.7% [ 2.6-19.0%] |
| | 3 | 0 | 0.0% | 2 [0-6] | 1.2% [0.0- 3.6%] | 0 | 0.0% | 1 [0-5] | 2.3% [ 0.0- 7.8%] |
| | 4 | 0 | 0.0% | 1 [0-4] | 0.6% [0.0- 2.4%] | 0 | 0.0% | 0 [0-3] | 0.0% [ 0.0- 4.3%] |
| | 5 | 0 | 0.0% | 0 [0-2] | 0.0% [0.0- 1.2%] | 0 | 0.0% | 0 [0-2] | 0.0% [ 0.0- 2.8%] |
| | 6 | 0 | 0.0% | 0 [0-2] | 0.0% [0.0- 1.2%] | 0 | 0.0% | 0 [0-1] | 0.0% [ 0.0- 1.9%] |
| | 7+ | 0 | 0.0% | 0 [0-2] | 0.0% [0.0- 1.2%] | 0 | 0.0% | 0 [0-1] | 0.0% [ 0.0- 2.1%] |
| | Total that grew | 15 | | 28 [15-43] | 63.1% [55.0-70.6%] | 26 | | 60 [43-86] | |
| | Total | 165 | | 165 [165-165] | 16.8% [12.0-22.3%] | 26 | | 60 [43-86] | |

† Parts of the actual Amsterdam transmission chains were observed in viral phylogenies of the major subtypes and circulating recombinant forms (B, 01AE, 02AG, C, D, G, A1 or 06cpx).
‡ Predicted based on the Bayesian branching process growth model and accounting for undiagnosed and unsampled individuals.

*Table 2: Growth distribution of Amsterdam transmission chains in 2014-2018.*

| | Observed Total (N) | In pre-existing chains (N) | (%) | In emerging chains (N) | (%) | Predicted Total (N) | In pre-existing chains (N) | (%) | In emerging chains (N) | (%) |
|---|---|---|---|---|---|---|---|---|---|---|
| MSM (Dutch) | 145 | 86 | 59.30% | 59 | 40.70% | 271 [216-340] | 145 [102-199] | 53.6% [44.2-62.3%] | 125 [100-158] | 46.4% [37.7-55.8%] |
| MSM (W. Europe, N. America, Oceania) | 40 | 25 | 62.50% | 15 | 37.50% | 73 [53-97] | 40 [25-59] | 54.5% [41.0-67.2%] | 33 [22-46] | 45.5% [32.8-59.0%] |
| MSM (E. & C. Europe) | 17 | 9 | 52.90% | 8 | 47.10% | 31 [20-45] | 16 [8-27] | 53.3% [33.3-71.9%] | 14 [7-23] | 46.7% [28.1-66.7%] |
| MSM (S. America & Caribbean) | 53 | 24 | 45.30% | 29 | 54.70% | 102 [77-133] | 54 [35-77] | 52.7% [40.4-64.1%] | 48 [34-65] | 47.3% [35.9-59.6%] |
| MSM (Other) | 42 | 14 | 33.30% | 28 | 66.70% | 81 [59-110] | 39 [24-59] | 48.1% [34.5-61.5%] | 42 [28-59] | 51.9% [38.5-65.5%] |
| **MSM (All)** | **297** | **158** | **53.20%** | **139** | **46.80%** | **559 [456-693]** | **294 [212-399]** | **52.8% [44.3-60.5%]** | **263 [217-323]** | **47.2% [39.5-55.7%]** |
| Heterosexual (Dutch) | 14 | 2 | 14.30% | 12 | 85.70% | 40 [24-61] | 15 [6-30] | 37.8% [18.2-58.5%] | 24 [14-39] | 62.2% [41.5-81.8%] |
| Heterosexual (Sub-Saharan Africa) | 11 | 4 | 36.40% | 7 | 63.60% | 32 [18-53] | 11 [3-26] | 34.6% [11.8-58.8%] | 21 [11-36] | 65.4% [41.2-88.2%] |
| Heterosexual (S. America & Caribbean) | 14 | 8 | 57.10% | 6 | 42.90% | 36 [21-61] | 15 [5-33] | 42.3% [18.2-65.9%] | 20 [11-36] | 57.7% [34.1-81.8%] |
| Heterosexual (Other) | 5 | 3 | 60.0% | 2 | 40.0% | 14 [6-24] | 5 [1-12] | 38.9% [10.0-70.6%] | 8 [3-15] | 61.1% [29.4-90.0%] |
| **Heterosexual (All)** | **44** | **17** | **38.60%** | **27** | **61.40%** | **122 [85-180]** | **46 [22-87]** | **38.7% [22.6-55.4%]** | **74 [51-110]** | **61.3% [44.6-77.4%]** |

† Parts of the actual Amsterdam transmission chains were observed in viral phylogenies of the major subtypes and circulating recombinant forms (B, 01AE, 02AG, C, D, G, A1 or 06cpx).
‡ Predicted based on the Bayesian branching process growth model and accounting for undiagnosed and unsampled individuals.

*Table 3: Amsterdam infections since 2014 in pre-2014 and emerging chains.*

### Emerging transmission chains outnumber pre-existing, growing transmission chains

We next used a Bayesian branching process growth model to predict the size and growth of the actual transmission chains, and account for the fact that larger proportions of recent infections remain undiagnosed and that approximately half of diagnosed individuals had a sequence sampled (see Materials and Methods, and Supplementary Material, Section 6). Model fit to the observed growth distributions was very good (Supplementary Figure S16).

We estimate that there are substantially more emerging chains in Amsterdam since 2014 than phylogenetically observed, 182 [161-207] in MSM and 60 [43-86] in heterosexuals, reflecting that emergent chains have a high probability to be entirely unobserved when growth is below the epidemic reproduction threshold of one (Table 2). Thus, the estimated actual, emerging chains outnumber the growing pre-2014 chains in both Amsterdam MSM and heterosexuals more strongly than the phylogenetic data suggest.

We estimate further that 61.3% [44.6-77.4%] of the estimated infections in 2014-2018 were in emerging chains among Amsterdam heterosexuals. However among Amsterdam MSM, 47.2% [39.5-55.7%] of the estimated infections in 2014-2018 were in emerging chains, suggesting circulating strains in Amsterdam prior to 2014 account for over half of new infections among MSM.

Proportion of locally preventable infections

From the emerging transmission chains, we can directly estimate the proportion of Amsterdam infections since 2014 that had an Amsterdam source (see Materials and Methods). We call these infections locally preventable, because they are within the reach of the HIV prevention efforts in Amsterdam. In Amsterdam MSM, an estimated 68% [61-74%] of infections in 2014-2018 were locally preventable, with little variation by region of birth (Figure 4A). In Amsterdam heterosexuals, an estimated 57% [41-71%] of infections in 2014-2018 were locally preventable, with more variation by region of birth, though we caution that the underlying sample sizes are small.

Figure 4B summarises the estimated number of locally preventable infections in Amsterdam in 2014-2018, obtained by multiplying the proportions of locally preventable infections with the estimated number of infections in 2014-2018 in each risk group. The majority of locally preventable Amsterdam infections in 2014-2018 were in foreign-born MSM (191 [139-262]), followed by Dutch-born MSM (184 [133-249]). See Supplementary Tables S3-S4 for details.

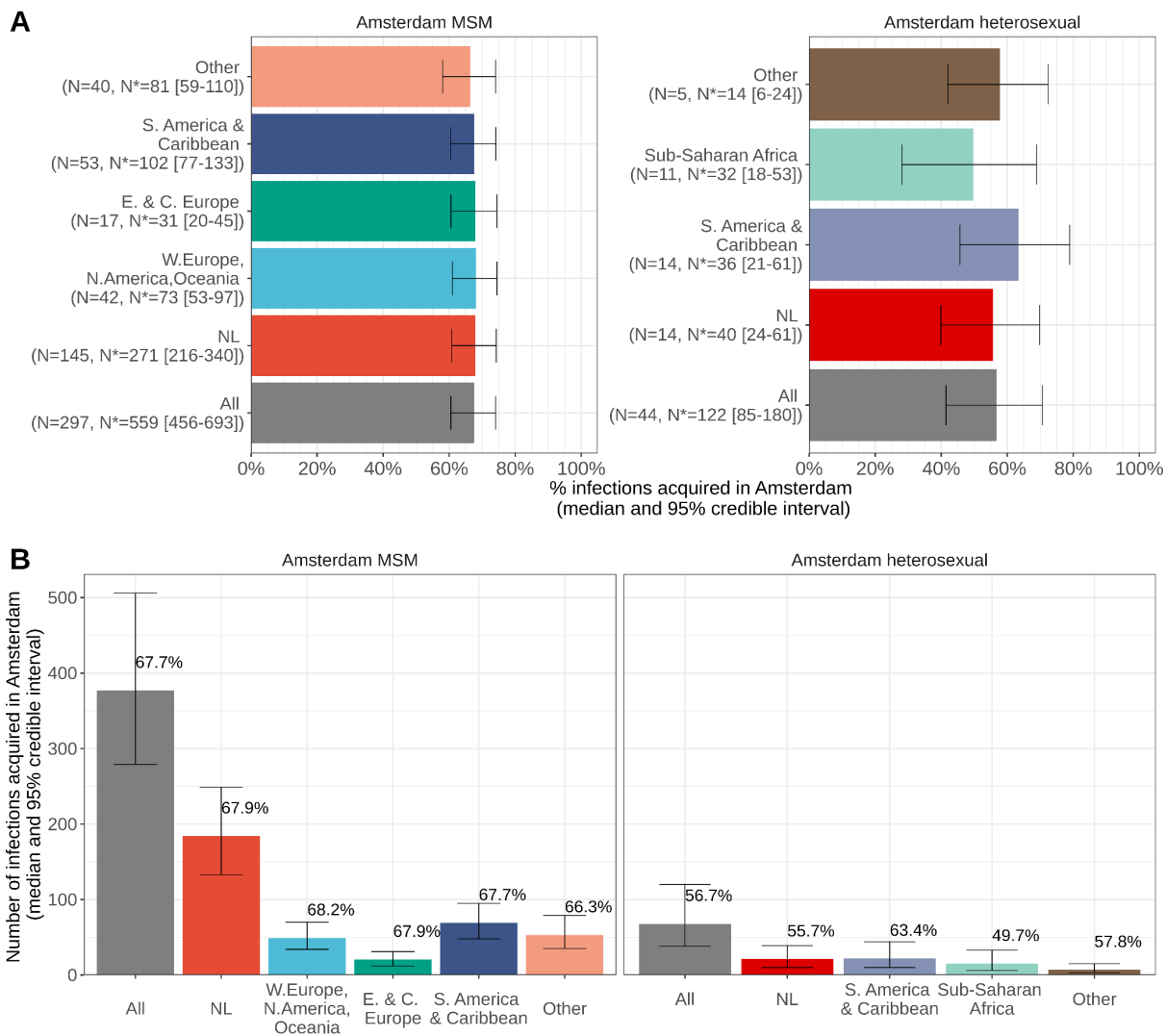

*Figure 4:* Estimated locally preventable infections in 2014-2018. (A) Posterior median estimates of the proportion of locally preventable infections along with 95% credible intervals, for MSM and heterosexuals stratified by place of birth. In the legend, N indicates the number of sequences available, and N* the estimated number of actual infections along with 95% credible intervals. (B) Estimated number of locally preventable infections along with 95% credible intervals, for MSM and heterosexuals stratified by place of birth. Posterior median estimates of proportion (%) of preventable infections shown above bars.

# Discussion

More than 300 cities have by the end of 2021 signed the Fast-Track Cities Paris Declaration and committed to end the AIDS epidemic by 2030, addressing disparities in access to basic health and social services, social justice and economic opportunities. The city of Amsterdam reached the UNAIDS Fast-track Cities 95-95-95 targets before the onset of the COVID-19 pandemic, and has seen a decade of declines in city-level HIV diagnoses. Here, we quantify the further potential of preventing HIV infection and recent annual spread at city level based on viral phylogenetic analysis of the cities' growing HIV transmission chains.

We can structure our insights in four themes. First, when focusing on the denominator of recent infections that are estimated to have occurred in 2014-2018, the proportion of undiagnosed individuals infected with HIV are at or above 20% in (self-identified) Amsterdam MSM risk groups, and at or above 30% in Amsterdam heterosexual risk groups. These results underscore that strategies aimed at raising awareness of HIV infection, providing easy access to checking symptoms of early HIV infection, encouraging frequent testing, PrEP provision, addressing fears of a positive test and reducing sitgma are vital to break the forefront of ongoing HIV transmission chains (H-TEAM Amsterdam n.d.; Maartje Dijkstra et al. 2017; Heijman et al. 2009; Burns, Rodger, and Johnson 2017; Myers et al. 2018). The estimated times to diagnosis document substantial disparities across risk groups in entering HIV care in Amsterdam, and separate efforts have characterised individuals with late diagnoses (Op de Coul et al. 2016; Bil et al. 2019; Slurink et al. 2021). One limitation of our approach is that time-to-diagnosis is estimated from clinical biomarkers and so, for individuals who recently arrived in Amsterdam, includes the potential time periods from infection to arrival. This prompted us to investigate potential upwards bias in our estimates of the proportion of undiagnosed Amsterdam infections, by comparing our estimates against those generated by the ECDC model for all the Netherlands. We found that the ECDC model resulted in larger undiagnosed proportions (Supplementary Material, Section 3.5), suggesting that our results are unlikely to be substantially upwards biased. We also validated time to infection estimates by comparing the estimated proportion of recent HIV infections ($\leq$6 months) with those estimated in an independent study in Amsterdam using avidity assays, and found them to be similar (Slurink et al. 2021). See Supplementary Figure S4 for details.

Second, we documented the growth of Amsterdam HIV transmission chains in which all phylogenetically observed members were virally suppressed by 2014. We find that regardless of risk group, almost all such virally suppressed chains did not grow in the sense that no new infections were phylogenetically observed. These results are unsurprising and mirror the established relationship that treatment for HIV infection, which results in undetectable viral load equals untransmittable virus. To track progress in tackling inequalities between those suppressed, and going the last mile to end the AIDS epidemic, we suggest that sequences are obtained from all newly diagnosed patients and the suppression status of transmission chains is routinely monitored.

Third, we initially speculated that with a decade of declining HIV diagnoses in Amsterdam, those infections that still occur might be concentrated in newly seeded, emerging transmission chains. It is challenging to interpret the directly observed data because high proportions of individuals remain undiagnosed and/or are not sequenced, and emerging chains are more likely to be completely undetected. We thus used statistical growth models accounting for unsampled cases, and we estimate in contrast to our initial speculations that 53% of new Amsterdam MSM infections in 2014-2018 grew from chains that existed prior to 2014, and 39% of new Amsterdam heterosexual infections. Following up and tracing back from known transmission chains is easier than discovering emerging chains, and so the many new infections that originate in existing chains have particularly high prevention potential (Oster, France, and Mermin 2018; Little et al. 2021; Dennis et al. 2021).

Fourth, we characterised the locally preventable Amsterdam infections in 2014-2018 by key population, i.e. MSM and heterosexuals stratified by region of birth. We defined locally preventable infections as the infections in Amsterdam residents in 2014-2018 who are estimated to have a source in another Amsterdam resident. The locally preventable infections thus comprise all new infections in pre-2014 chains and all new infections in emerging chains except their index case. We estimate that 68% of infections in Amsterdam MSM were locally preventable, and 57% in Amsterdam heterosexuals. One limitation of our analyses is that in addition to undiagnosed individuals, viral sequences were missing for 33% of diagnosed MSM and for 37% of diagnosed heterosexuals. We thus found data on fewer of the actual transmission chains than planned, and our sampling frame is not powered to identify statistically meaningful differences in the proportion of locally preventable infections by risk group. More detailed estimates into the locally preventable infections among migrants have recently been obtained through clinic surveys across Europe (Arco et al. 2017). These data suggest similar estimates of in-country HIV acquisition post migration of 51% in heterosexual women and 58% in heterosexual men, which are consistent with ours, and further highlight important variations by place of birth and other demographic characteristics that we are unable to uncover.

In summary, we find considerable potential to prevent Amsterdam HIV infections, which could be targeted through city-level interventions, even in the context of substantial improvements in curbing the number of diagnoses and infections in Amsterdam over the past 10 years. Given the similarity of the demographics, HIV burden, access to care, and prevention approaches across many cities in Western Europe, our conclusions are relevant to many UNAIDS Fast-Track cities, and provide evidence-based support for locally-targeted combination HIV prevention interventions in metropolitan areas. In the meantime COVID-19 is severely disrupting prevention messaging, testing and PrEP services, and early pathways to care, making innovative and targeted prevention approaches all the more important.

## Acknowledgments

We thank the steering committee of the Amsterdam HIV transmission initiative for earlier comments on this work; and Imperial College Research Computing Service, DOI: 10.14469/h-pc/2232, for providing the computational resources to perform this study.

This work received funding as part of the H-TEAM initiative from Aidsfonds (project number P29701). The H-TEAM initiative is being supported by Aidsfonds (grant number: 2013169, P29701, P60803), Stichting Amsterdam Dinner Foundation, Bristol-Myers Squibb International Corp. (study number: AI424-541), Gilead Sciences Europe Ltd (grant number: PA-HIV-PREP-16-0024), Gilead Sciences (protocol numbers: CO-NL-276-4222, CO-US-276-1712, CO-NL-985-6195), and M.A.C AIDS Fund.

## Author contributions

OR conceived the study. OR, GdB, PR, CF, MP, TvdL, EOdC, CF, DB oversaw the study. AvS, DB prepared the data. AB, OR developed the branching process model. AB, MM, NP performed all analyses. AB, OR wrote the draft manuscript. All authors contributed to and approved the final manuscript.

# Data availability

HIV physicians can review the data of their own treatment centre and compare these data with the full cohort through an online report builder. Statistical information or data for own research purposes can be requested by submitting a research proposal (https://www.hiv-monitoring.nl/english/research/research-projects/). For correspondence: hiv.monitoring@amc.uva.nl.

# Code availability

Code is available on Github, https://github.com/alexblenkinsop/bpm.

# Supplementary Material

Supplementary Text
Supplementary Tables S1-S5
Supplementary Figures S1-S28

Supplementary Text to

# Estimating the potential to prevent locally acquired HIV infections in a UNAIDS Fast-Track City, Amsterdam

Blenkinsop et. al.

# Contents

# S1 Supplementary Tables and Figures

| Strata | | All patients | Patients with a sequence† |
|---|---|---|---|
| Sex | Female | 40 (7.8%) | 24 (7%) |
| | Male | 476 (92.2%) | 317 (93%) |
| Risk group | MSM | 446 (86.4%) | 297 (87.1%) |
| | Heterosexual | 70 (13.6%) | 44 (12.9%) |
| Age group at estimated time of infection | 18-24 | 74 (14.3%) | 48 (14.1%) |
| | 25-34 | 209 (40.5%) | 124 (36.4%) |
| | 35-44 | 113 (21.9%) | 76 (22.3%) |
| | 45-59 | 110 (21.3%) | 87 (25.5%) |
| | 60+ | 10 (1.9%) | 6 (1.8%) |
| Place of birth | Sub-Saharan Africa | 24 (4.8%) | 16 (4.8%) |
| | Asia | 20 (4%) | 13 (3.9%) |
| | Australia & New Zealand | 2 (0.4%) | 2 (0.6%) |
| | Central Europe | 25 (5%) | 16 (4.8%) |
| | Eastern Europe | 8 (1.6%) | 1 (0.3%) |
| | Suriname, Curacao & Aruba | 41 (8.1%) | 32 (9.6%) |
| | South America & Caribbean | 63 (12.5%) | 35 (10.5%) |
| | Middle East & North Africa | 31 (6.1%) | 20 (6%) |
| | Netherlands | 213 (42.2%) | 159 (47.6%) |
| | North America | 23 (4.6%) | 14 (4.2%) |
| | Western Europe | 55 (10.9%) | 26 (7.8%) |
| Estimated time to diagnosis (years) | | 0.4 [0.04-3.2] | 0.41 [0.03-3.25] |

† Patients with sequence of a subtype or circulating recombinant form B, 01AE, 02AG, C, D, G, A1 or 06cpx.

Table S1: Patient characteristics for individuals with an estimated infection date between 2014-2018.

| Risk group | Subtype | All sampled individuals virally suppressed by 2014† | Pre-2014 chains | | Pre-2014 chains that grew | | Individuals (Total) | Individuals (infected before 2014) | | Individuals (infected before 2014 & not virally suppressed | | Individuals (infected 2014-2018) | |
|---|---|---|---|---|---|---|---|---|---|---|---|---|---|
| | | | (n) | (n) | (%) | | (n) | (n) | (%) | (n) | (%) | (n) | (%) |
| Amsterdam MSM | B | Yes | 866 | 35 | 4% | | 1432 | 1279 | 89% | 0 | 0% | 153 | 11% |
| | B | No | 286 | 44 | 15% | | 1740 | 1303 | 75% | 352 | 20% | 85 | 5% |
| | Non-B | Yes | 83 | 8 | 10% | | 172 | 119 | 69% | 0 | 0% | 53 | 31% |
| | Non-B | No | 18 | 2 | 11% | | 80 | 51 | 64% | 23 | 29% | 6 | 8% |
| | Total | | 1253 | 89 | 7% | | 3424 | 2752 | 80% | 375 | 11% | 297 | 9% |
| AmsHSX | B | Yes | 180 | 5 | 3% | | 218 | 200 | 92% | 0 | 0% | 18 | 8% |
| | B | No | 85 | 4 | 5% | | 284 | 189 | 67% | 90 | 32% | 5 | 2% |
| | Non-B | Yes | 221 | 5 | 2% | | 301 | 281 | 93% | 0 | 0% | 20 | 7% |
| | Non-B | No | 90 | 1 | 1% | | 235 | 142 | 60% | 92 | 39% | 1 | 0% |
| | Total | | 576 | 15 | 3% | | 1038 | 812 | 78% | 182 | 18% | 44 | 4% |
| | Total | | 1829 | 104 | 6% | | 4462 | 3564 | 80% | 557 | 12% | 341 | 8% |

† Individuals infected prior to 2014, with last viral load measurement before 2014 below 100copies/ml.

Table S2: Viral suppression status of the phylogenetically observed pre-2014 Amsterdam transmission chains.

| Transmission group | Birth place | Proportion of infections acquired in Amsterdam 2014-2018 | Number of infections acquired in Amsterdam 2014-2018 |
|---|---|---|---|
| Amsterdam MSM | All | 67.7% [60.5-74.1%] | 377 [279-506] |
| Amsterdam MSM | NL | 67.9% [60.8-74.3%] | 184 [133-249] |
| Amsterdam MSM | W.Europe, N.America, Oceania | 68.2% [61-74.5%] | 49 [33.975-70] |
| Amsterdam MSM | E. & C. Europe | 67.9% [60.6-74.5%] | 20.5 [12-31] |
| Amsterdam MSM | S. America & Caribbean | 67.7% [60.5-74.2%] | 69 [48-95] |
| Amsterdam MSM | Other | 66.3% [58-74.1%] | 53 [35-79] |
| Amsterdam heterosexual | All | 56.7% [41.5-70.6%] | 68 [38-120] |
| Amsterdam heterosexual | NL | 55.7% [39.9-69.8%] | 21 [10-39] |
| Amsterdam heterosexual | S. America & Caribbean | 63.4% [45.7-78.9%] | 22 [10-44] |
| Amsterdam heterosexual | Sub-Saharan Africa | 49.7% [28.1-68.9%] | 15 [6-33] |
| Amsterdam heterosexual | Other | 57.8% [42-72.4%] | 7 [3-15] |

Table S3: Estimated total number of infections acquired locally from an Amsterdam source, by transmission group and place of birth.

| Transmission group | Birth place | Number of infections acquired in Amsterdam 2014-2018 |
|---|---|---|
| Amsterdam MSM | Dutch-born | 184 [133-249] |
| Amsterdam MSM | Foreign-born | 191 [139-262] |
| Amsterdam heterosexual | Dutch-born | 21 [10-39] |
| Amsterdam heterosexual | Foreign-born | 45 [24-82] |

Table S4: Estimated total number of infections acquired locally from an Amsterdam source, by transmission group and whether individuals are Dutch-born or foreign-born.

| Risk group | Subtype | Origin of chains | Observed (N) | Observed (%) | Predicted (N) | Predicted (%) |
|---|---|---|---|---|---|---|
| Amsterdam MSM | B | Amsterdam - other risk group | 1 [1-3] | 0.8% [0.5-2%] | 2 [1-6] | 0.5% [0.2-1.4%] |
| | | Asia | 2 [2-4] | 1.5% [1-2.3%] | 6 [2-12] | 1.5% [0.5-2.9%] |
| | | Eastern Europe & Central Asia | 7 [4-13] | 5% [2.9-7.3%] | 21 [13-31] | 5% [3.1-7.3%] |
| | | South America & Caribbean | 5 [2-12] | 3.2% [1.5-5.9%] | 14 [8-22] | 3.4% [1.8-5.3%] |
| | | Middle East & North Africa | 1 [1-2] | 0.8% [0.5-1.3%] | 3 [1-7] | 0.7% [0.2-1.7%] |
| | | Netherlands | 96 [84-159] | 71.1% [64-77.1%] | 299 [277-323] | 71.2% [66.7-75.4%] |
| | | North America | 8 [4-17] | 5.7% [2.5-9.3%] | 24 [15-34] | 5.6% [3.5-7.9%] |
| | | Oceania | 2 [2-2] | 1% [1-1%] | 1 [1-2] | 0.2% [0.2-0.5%] |
| | | Western Europe | 16 [11-29] | 11.7% [8-15.9%] | 49 [37-63] | 11.7% [8.7-14.9%] |
| | Non-B | Sub-Saharan Africa | 3 [1-5] | 10.7% [3.6-19.6%] | 8 [3-13] | 10.7% [4.3-18.5%] |
| | | Amsterdam - other risk group | 1 [1-3] | 3.9% [3.3-11.4%] | 2 [1-4] | 2.5% [1.3-5.8%] |
| | | Asia | 8 [6-11] | 31% [22.2-42.3%] | 22 [14-30] | 31.4% [21-42.2%] |
| | | Eastern Europe & Central Asia | 1 [1-1] | 3.5% [3.3-3.6%] | 1 [1-2] | 1.4% [1.2-2.4%] |
| | | South America & Caribbean | 1 [1-2] | 4% [3.3-8.2%] | 3 [1-7] | 4.3% [1.4-9.8%] |
| | | Middle East & North Africa | 1 [1-1] | 3.6% [3.3-4%] | 1 [1-3] | 1.5% [1.2-4.1%] |
| | | Netherlands | 12 [8-16] | 46.4% [32.1-59.5%] | 32 [23-42] | 46% [34.4-57.8%] |
| Amsterdam heterosexual | B | Amsterdam - other risk group | 3 [1-7] | 21.4% [7.4-38.5%] | 22 [14-31] | 21.4% [13.5-29.7%] |
| | | Eastern Europe & Central Asia | 1 [1-1] | 7.2% [6.7-7.7%] | 1 [1-2] | 1% [0.8-2%] |
| | | Netherlands | 11 [8-17] | 75% [54.8-92%] | 76 [64-90] | 74.8% [66-83%] |
| | | North America | 1 [1-3] | 6.7% [4.7-10.6%] | 2 [1-4] | 1.9% [0.9-4.3%] |
| | | Western Europe | 1 [1-3] | 7.1% [5.3-20.3%] | 2 [1-6] | 2.1% [0.9-5.5%] |
| | Non-B | Sub-Saharan Africa | 5 [2-8] | 33.3% [9.4-51.9%] | 40 [29-52] | 32% [24-40.5%] |
| | | Amsterdam - other risk group | 1 [1-2] | 6.7% [5.4-12.5%] | 9 [4-15] | 6.9% [3-11.7%] |
| | | Asia | 1 [1-1] | 6.7% [5.7-9.8%] | 2 [1-6] | 1.7% [0.8-4.6%] |
| | | Netherlands | 8 [4-12] | 50% [28.9-74.2%] | 63 [50-76] | 50.4% [41.9-59.4%] |
| | | North America | 1 [1-1] | 5.6% [5.6-5.6%] | 1 [1-2] | 0.8% [0.7-1.6%] |

Table S5: Observed and estimated ancestral origins of phylogenetic subgraphs and complete transmission chains.

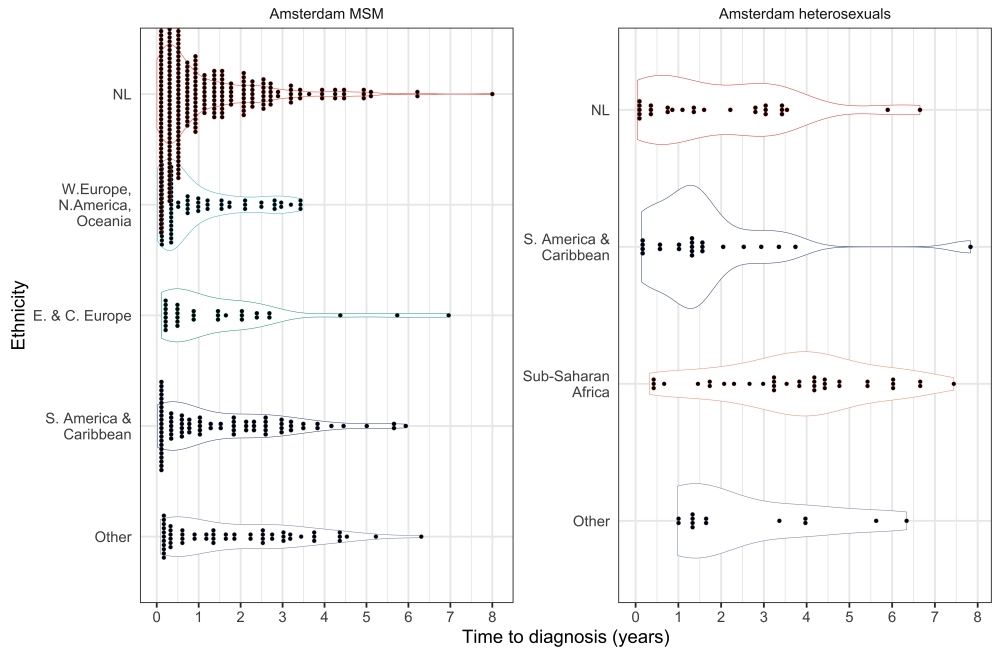

Figure S1: Distribution of individual level posterior median estimated times to diagnosis by place of birth, for Amsterdam MSM and heterosexuals.

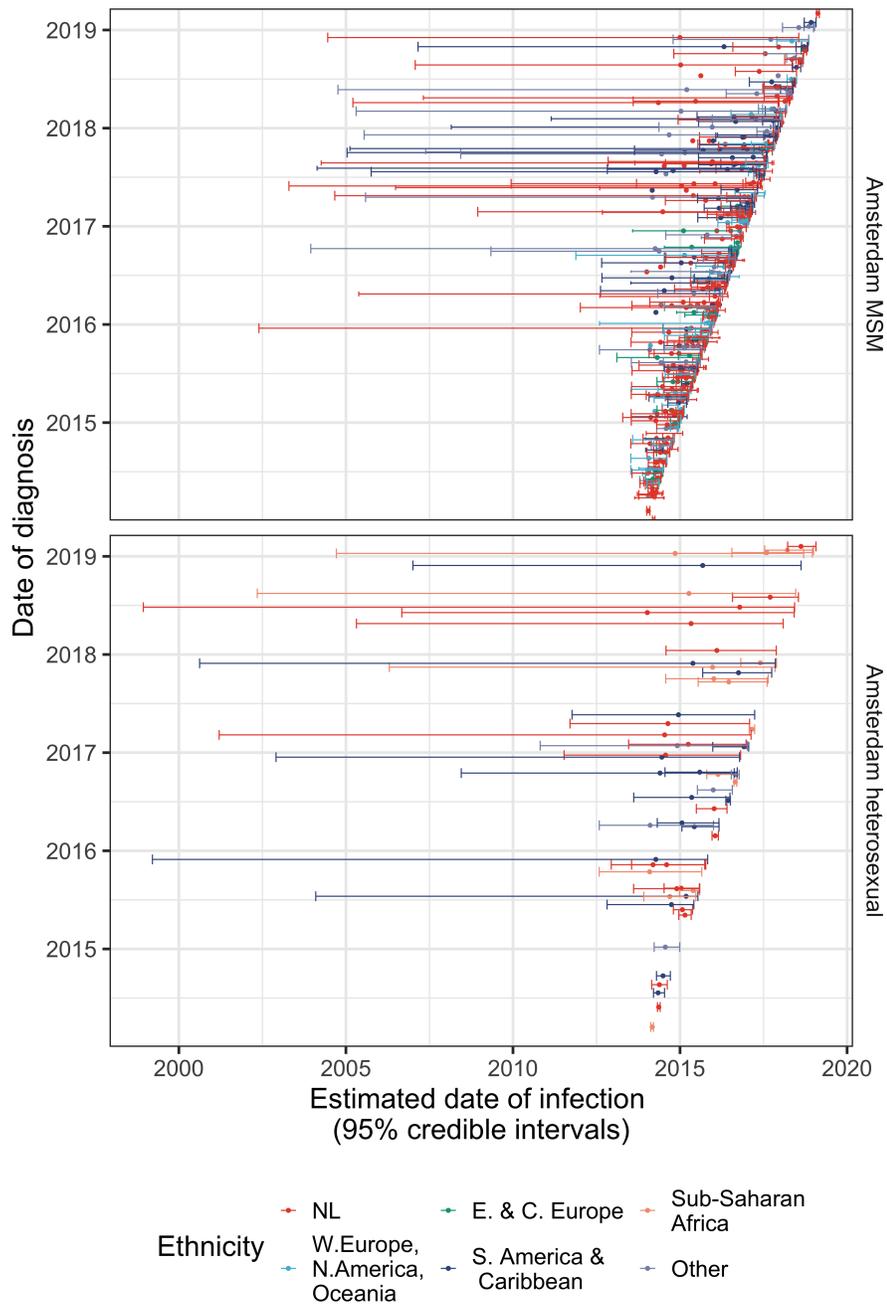

Figure S2: Diagnosis date and posterior median estimated infection date (with 95% credible interval) of individuals in Amsterdam diagnosed between January 2014 and May 2019.

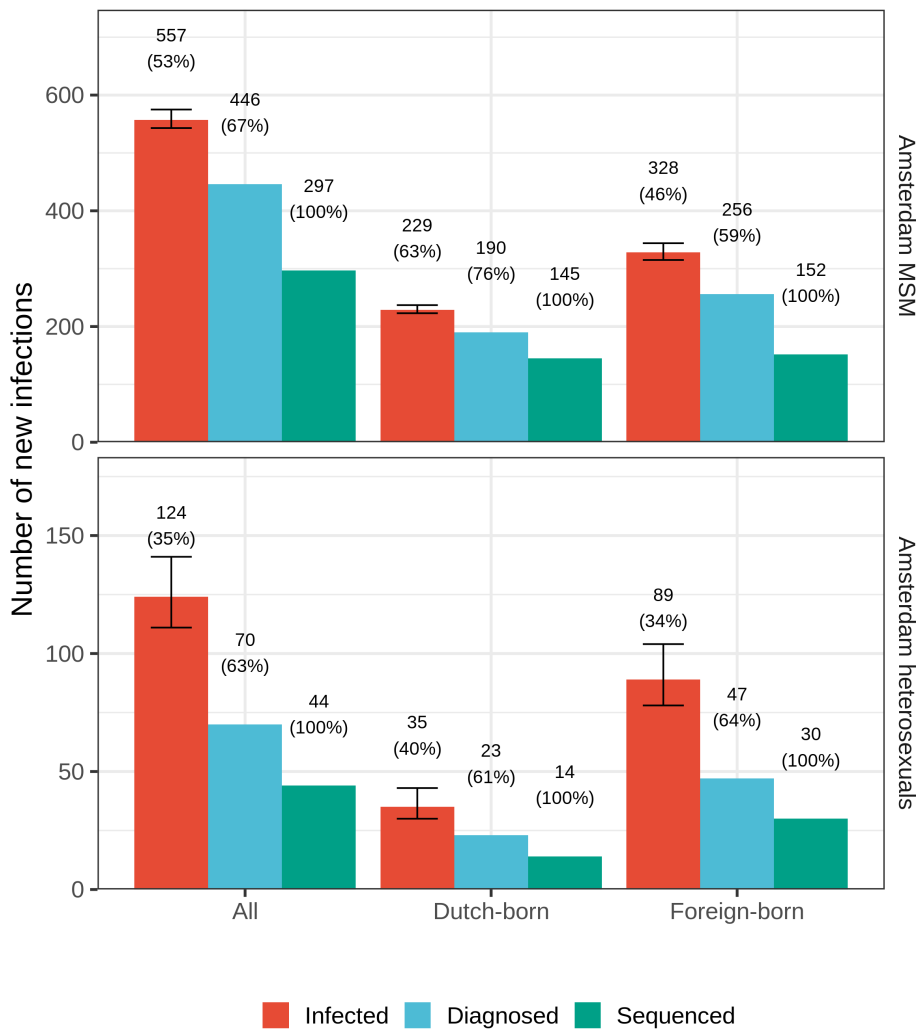

Figure S3: Estimated Amsterdam infections in 2014-2018. Estimates of the total number of individuals resident in Amsterdam that were infected in 2014-2018 are shown along with the subset of individuals that were diagnosed, and the subset of those for who at least one viral sequence is available. Posterior median estimates (bars, and number on top of bar) are shown along with 95% credible intervals. The posterior median proportion of individuals with a viral sequence is also shown (proportion on top of bar).

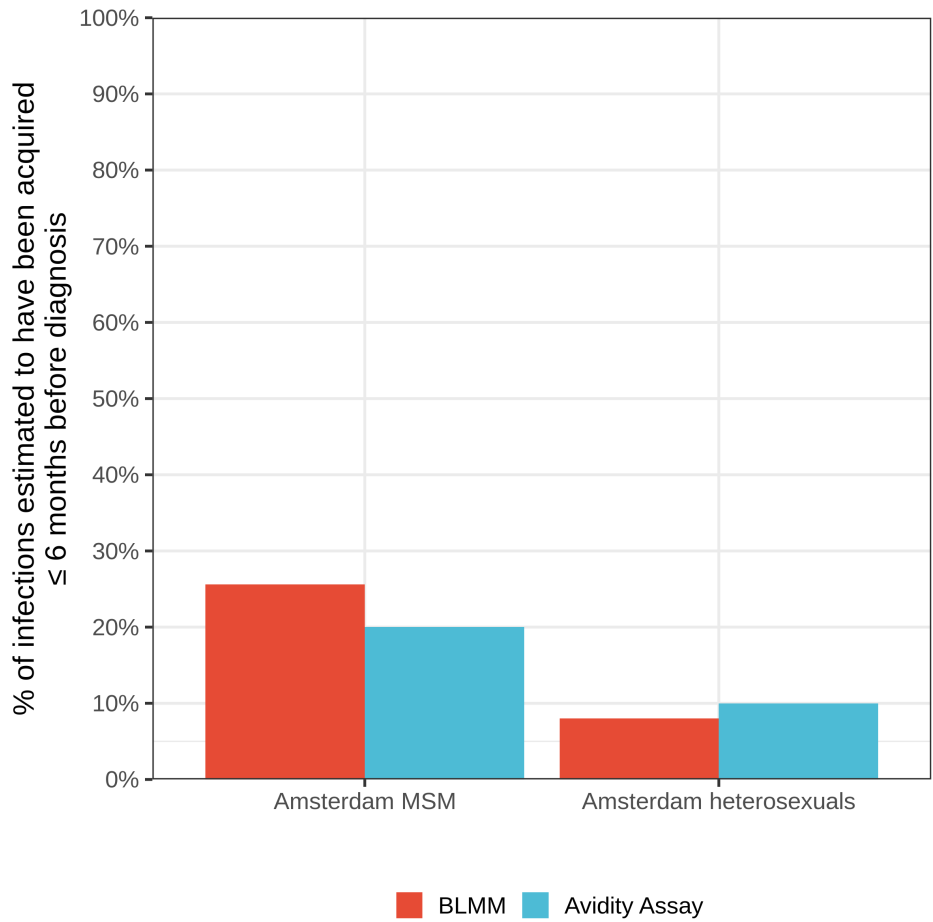

Figure S4: Estimates for the proportion of HIV infections acquired within 6 months of diagnosis from the bivariate linear mixed model (BLMM) approach (for infections diagnosed between 2013-2015), compared with estimates obtained from avidity assays in a study by Slurink et al.[1] (for infections diagnosed between 2013-2015).

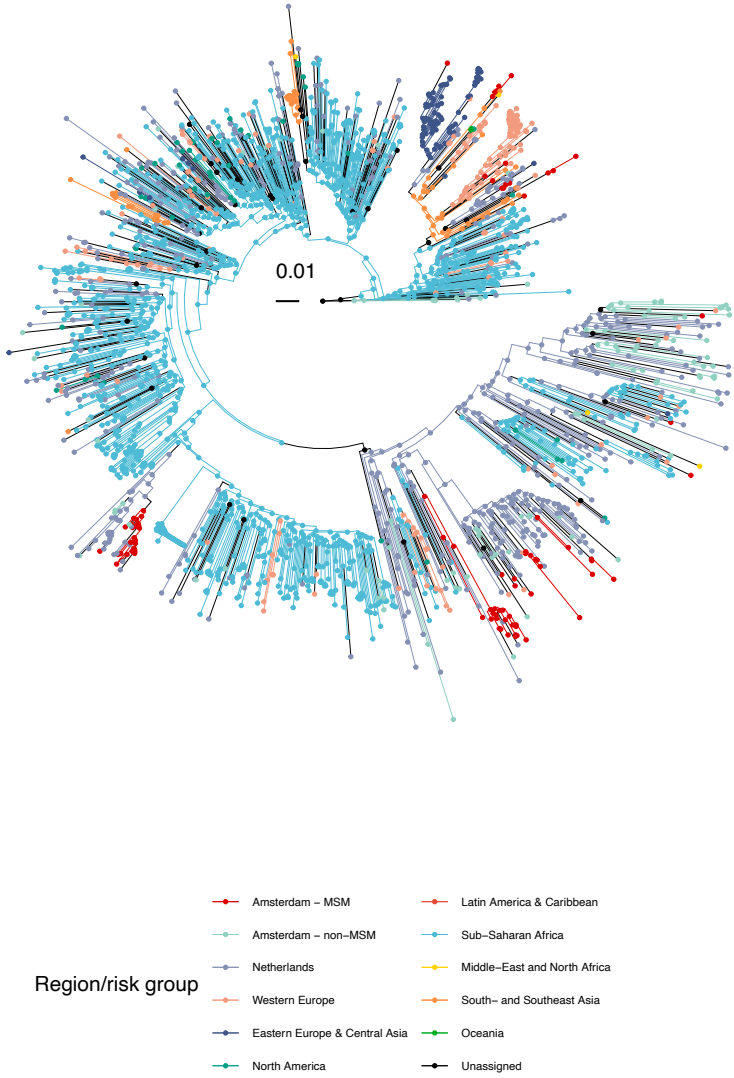

Figure S5: Annotated phylogeny of viral sequences of circulating recombinant form A1 of Amsterdam MSM and background individuals. Colours of tips show the observed states of each observed sequence, and colours of lineages represent inferred states. States were assigned to each sequence as described in equations S21-S22, and represent both transmission group (MSM, heterosexual, other) and place of birth or residence.

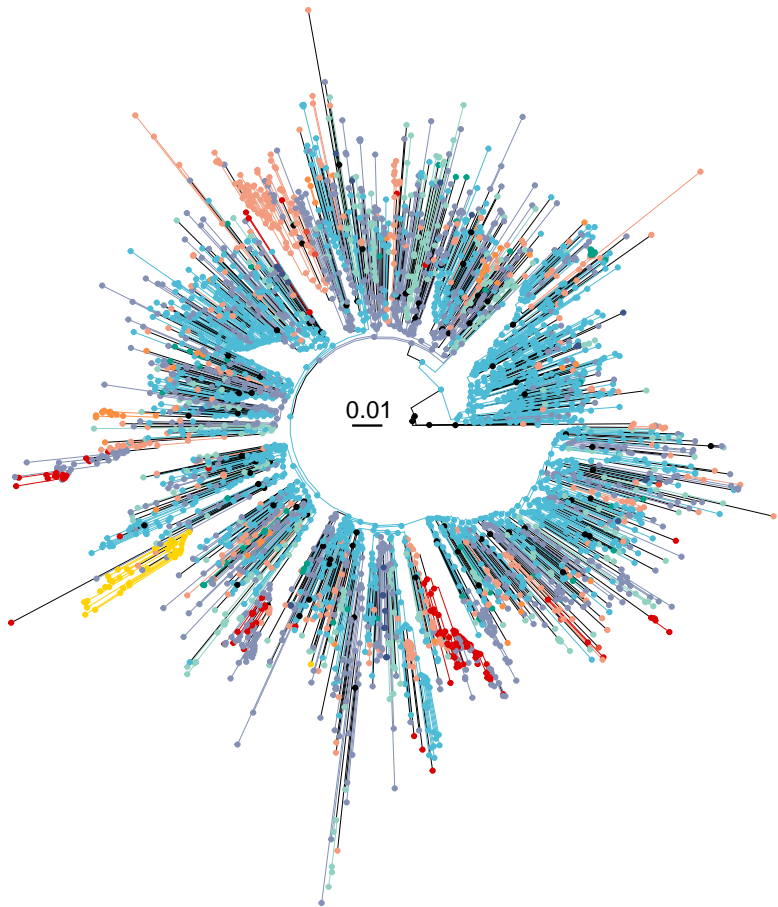
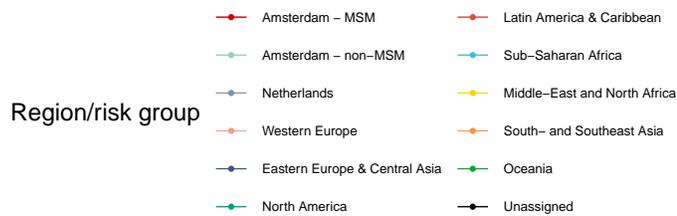

Figure S6: Annotated phylogeny of viral sequences of circulating recombinant form 02AG of Amsterdam MSM and background individuals. Colours of tips show the observed states of each observed sequence, and colours of lineages represent inferred states. States were assigned to each sequence as described in equations S21-S22, and represent both transmission group (MSM, heterosexual, other) and place of birth or residence.

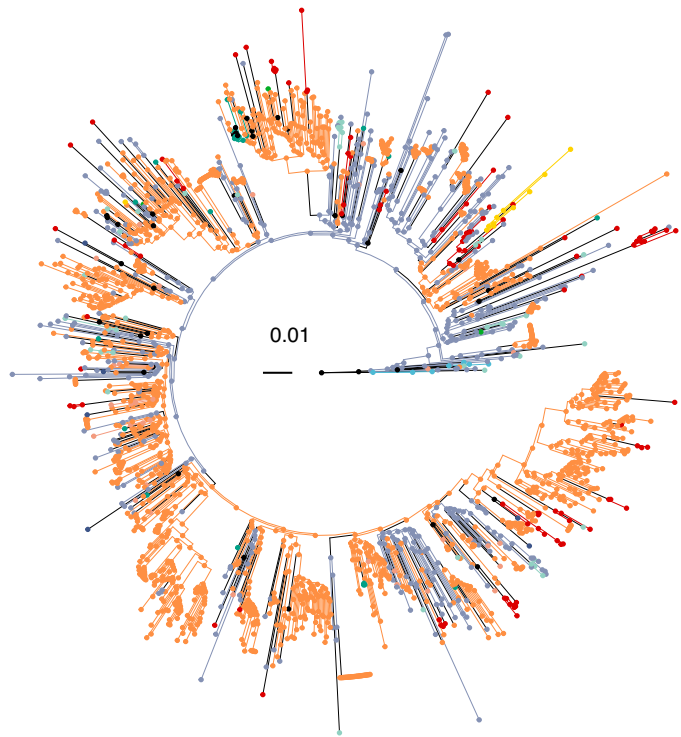

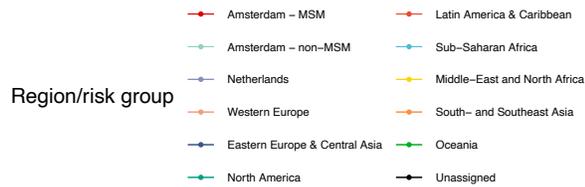

Figure S7: Annotated phylogeny of viral sequences of circulating recombinant form 01AE of Amsterdam MSM and background individuals. Colours of tips show the observed states of each observed sequence, and colours of lineages represent inferred states. States were assigned to each sequence as described in equations S21-S22, and represent both transmission group (MSM, heterosexual, other) and place of birth or residence.

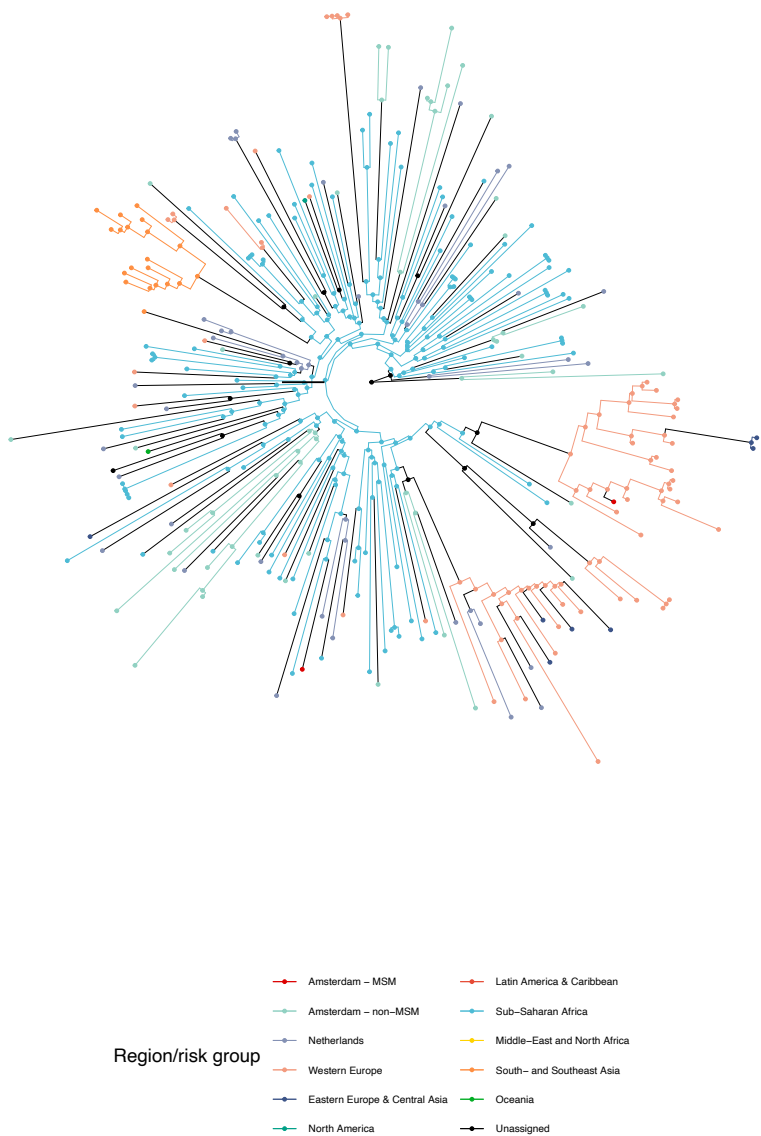

Figure S8: Annotated phylogeny of viral sequences of circulating recombinant form 06cpx of Amsterdam MSM and background individuals. Colours of tips show the observed states of each observed sequence, and colours of lineages represent inferred states. States were assigned to each sequence as described in equations S21-S22, and represent both transmission group (MSM, heterosexual, other) and place of birth or residence.

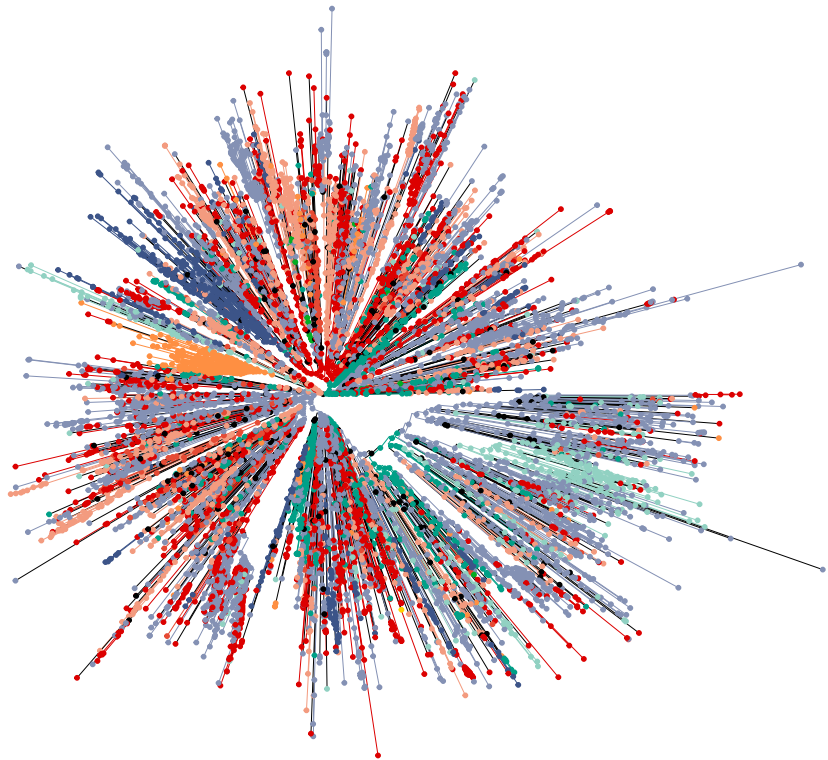

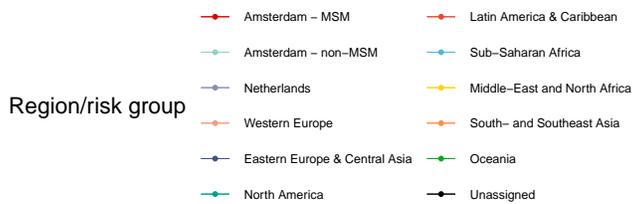

Figure S9: Annotated phylogeny of viral sequences of a sub-clade of subtype B of Amsterdam MSM and background individuals. Colours of tips show the observed states of each observed sequence, and colours of lineages represent inferred states. States were assigned to each sequence as described in equations S21-S22, and represent both transmission group (MSM, heterosexual, other) and place of birth or residence.

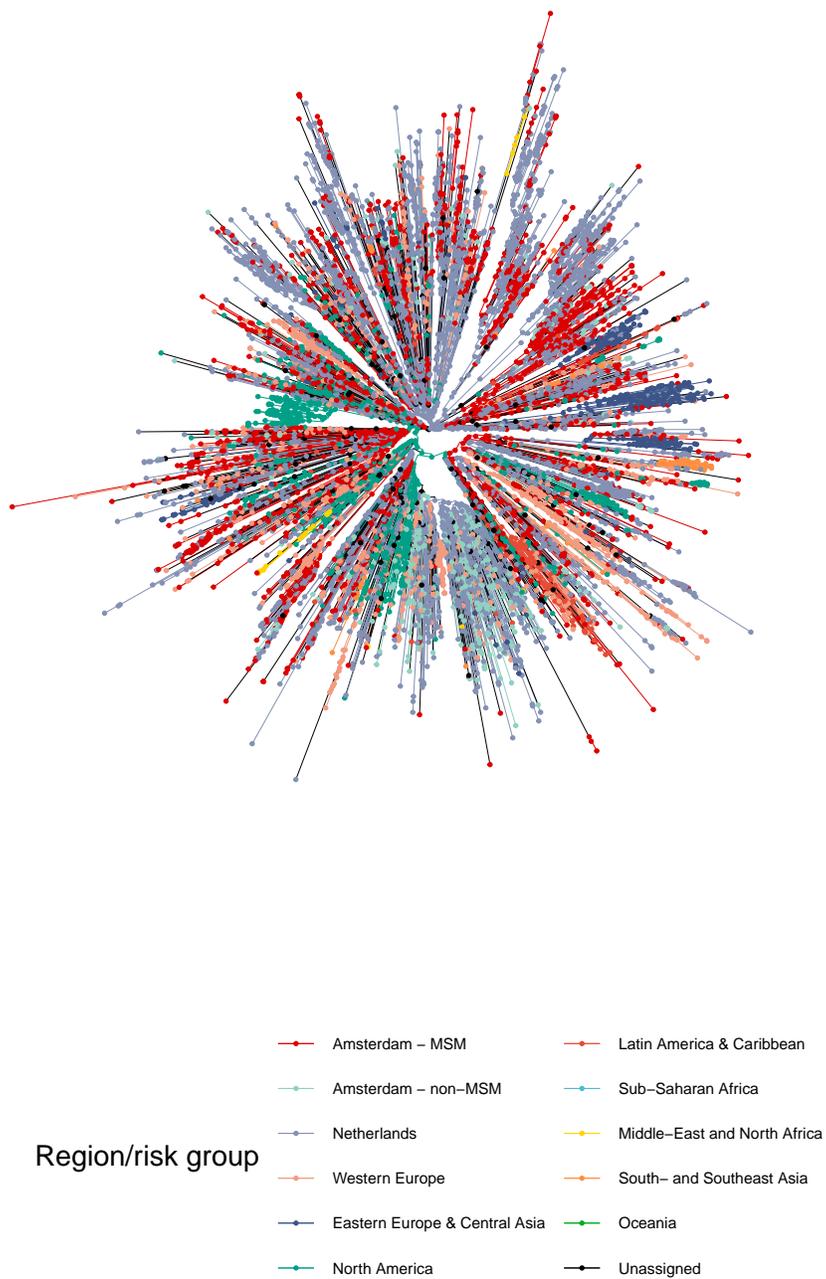

Figure S10: Annotated phylogeny of viral sequences of a sub-clade of subtype B of Amsterdam MSM and background individuals. Colours of tips show the observed states of each observed sequence, and colours of lineages represent inferred states. States were assigned to each sequence as described in equations S21-S22, and represent both transmission group (MSM, heterosexual, other) and place of birth or residence.

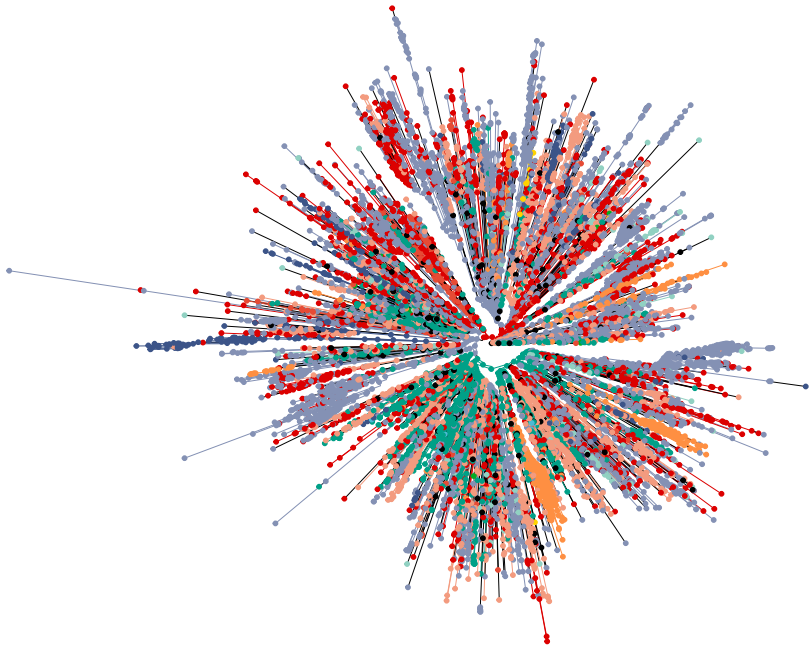

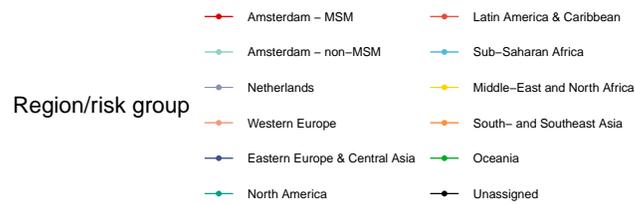

Figure S11: Annotated phylogeny of viral sequences of a sub-clade of subtype B of Amsterdam MSM and background individuals. Colours of tips show the observed states of each observed sequence, and colours of lineages represent inferred states. States were assigned to each sequence as described in equations S21-S22, and represent both transmission group (MSM, heterosexual, other) and place of birth or residence.

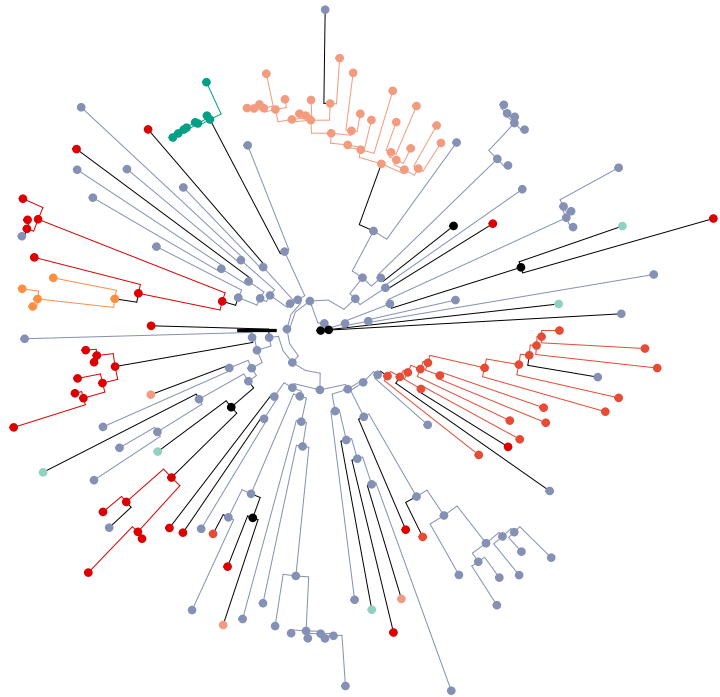

Figure S12: Annotated phylogeny of viral sequences of a sub-clade of subtype B of Amsterdam MSM and background individuals. Colours of tips show the observed states of each observed sequence, and colours of lineages represent inferred states. States were assigned to each sequence as described in equations S21-S22, and represent both transmission group (MSM, heterosexual, other) and place of birth or residence.

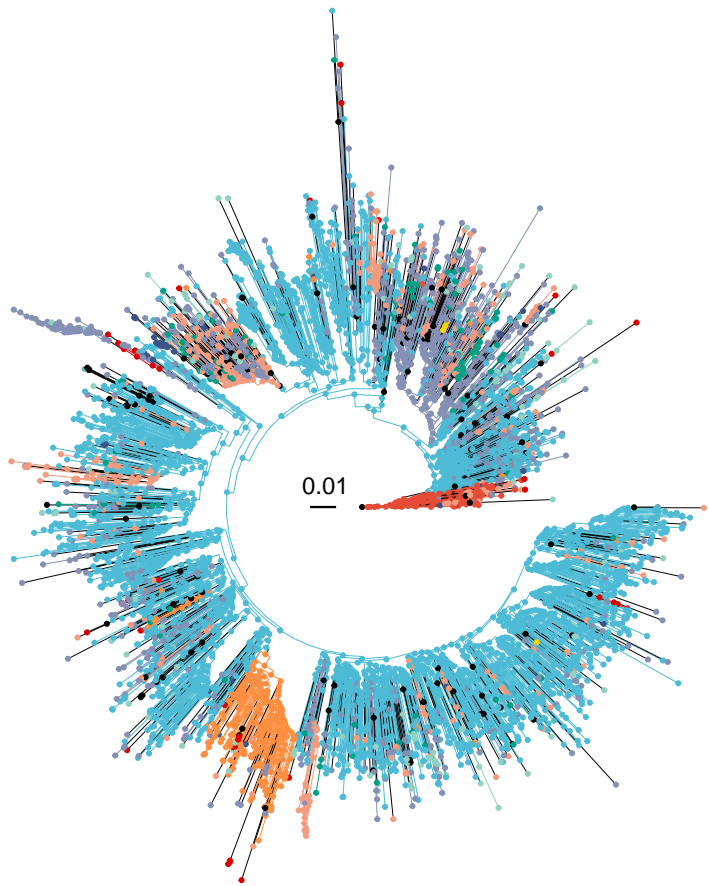
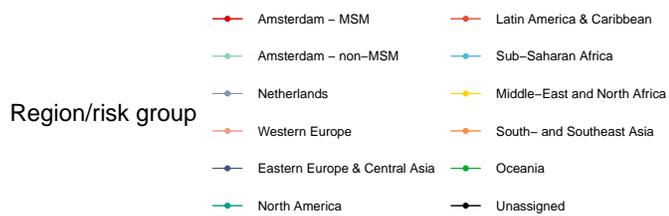

Figure S13: Annotated phylogeny of viral sequences of subtype C of Amsterdam MSM and background individuals. Colours of tips show the observed states of each observed sequence, and colours of lineages represent inferred states. States were assigned to each sequence as described in equations S21-S22, and represent both transmission group (MSM, heterosexual, other) and place of birth or residence.

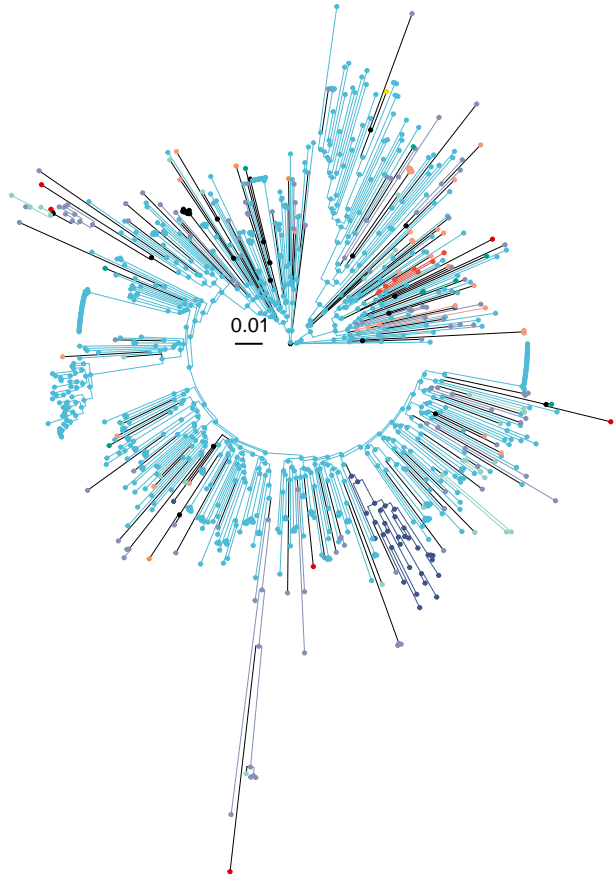
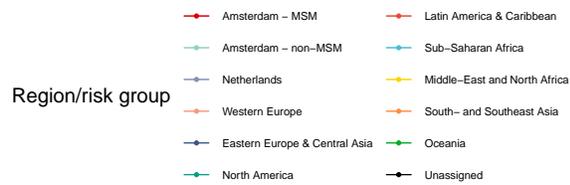

Figure S14: Annotated phylogeny of viral sequences of subtype D of Amsterdam MSM and background individuals. Colours of tips show the observed states of each observed sequence, and colours of lineages represent inferred states. States were assigned to each sequence as described in equations S21-S22, and represent both transmission group (MSM, heterosexual, other) and place of birth or residence.

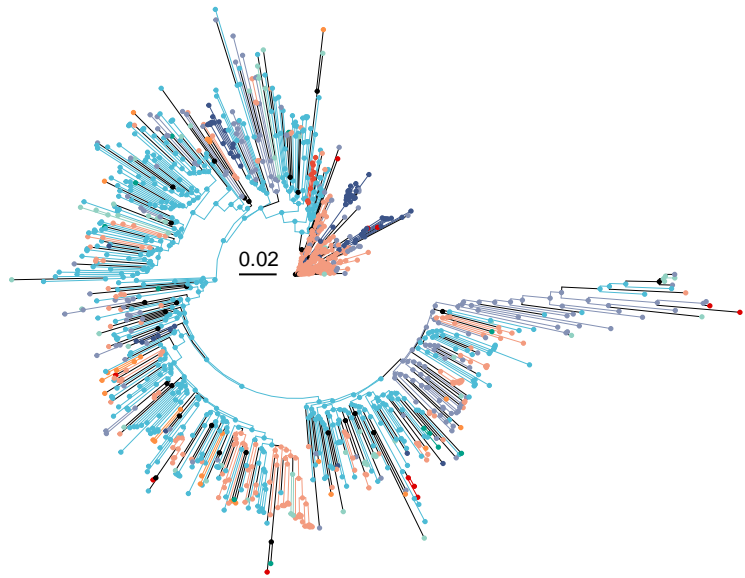

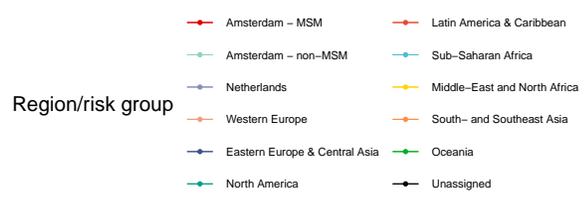

Figure S15: Annotated phylogeny of viral sequences of subtype G of Amsterdam MSM and background individuals. Colours of tips show the observed states of each observed sequence, and colours of lineages represent inferred states. States were assigned to each sequence as described in equations S21-S22, and represent both transmission group (MSM, heterosexual, other) and place of birth or residence.

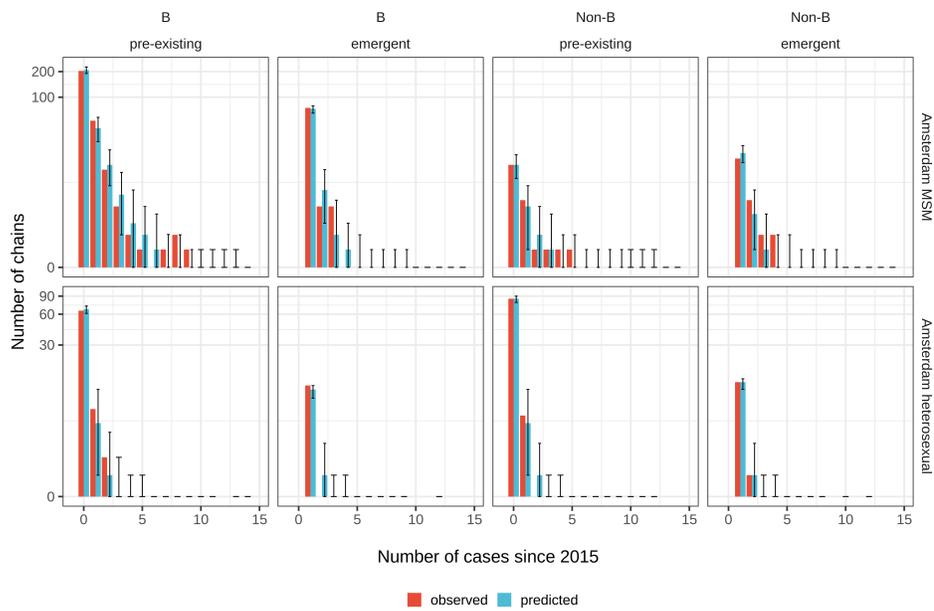

Figure S16: Posterior predictive check for Amsterdam MSM (top) and Amsterdam heterosexuals (bottom) for B and non-B subtypes.

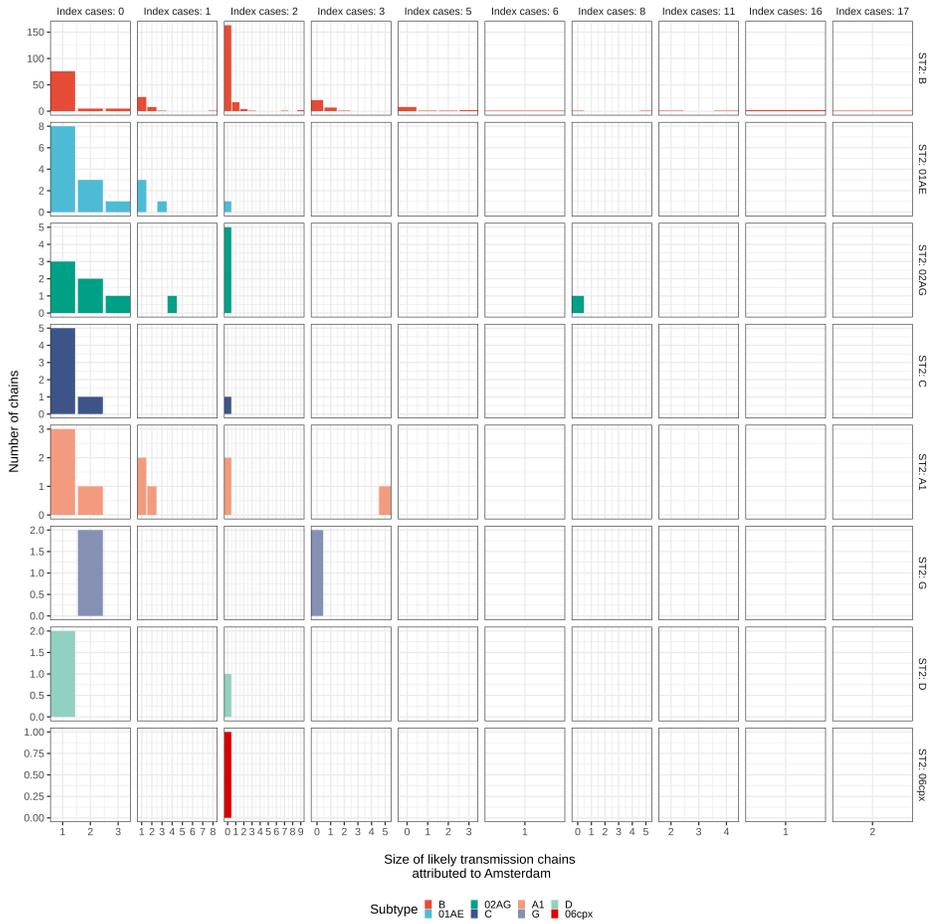

(a) Phylogenetic subgraphs among Amsterdam MSM.

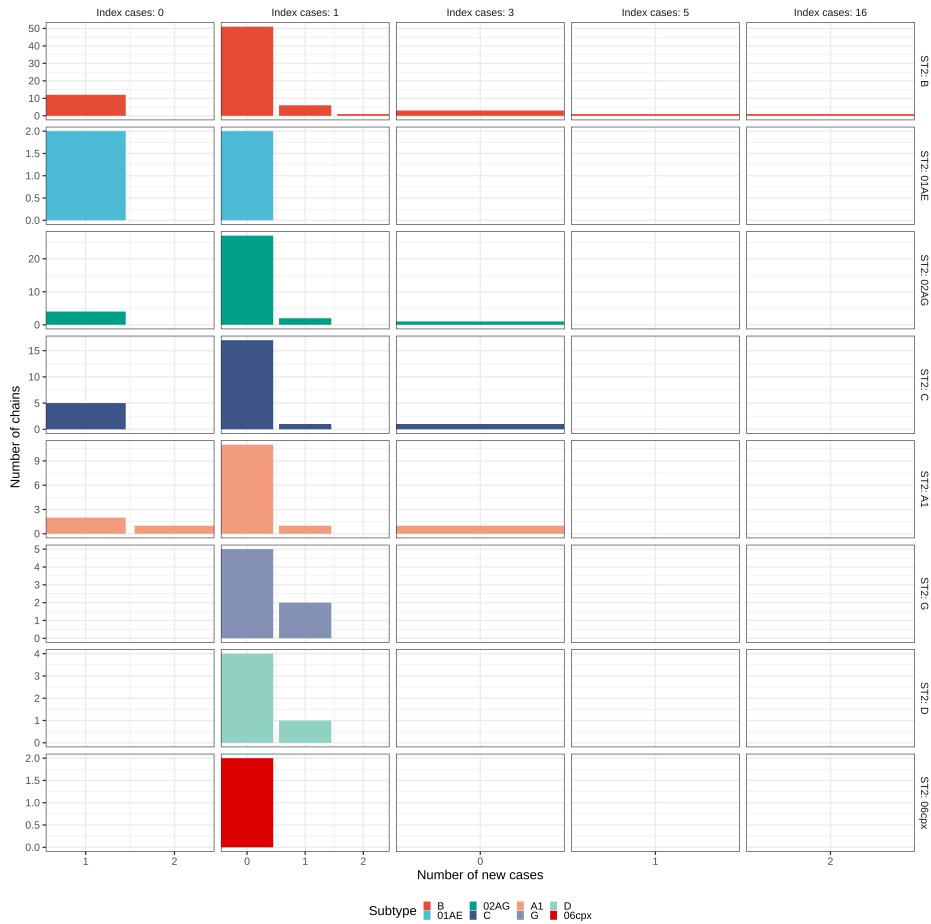

(b) Phylogenetic subgraphs among Amsterdam heterosexuals.

Figure S17: Growth of phylogenetically observed subgraphs by subtype. First column (index cases = 0 are for emergent chains, where the index case is among the newly generated cases.)

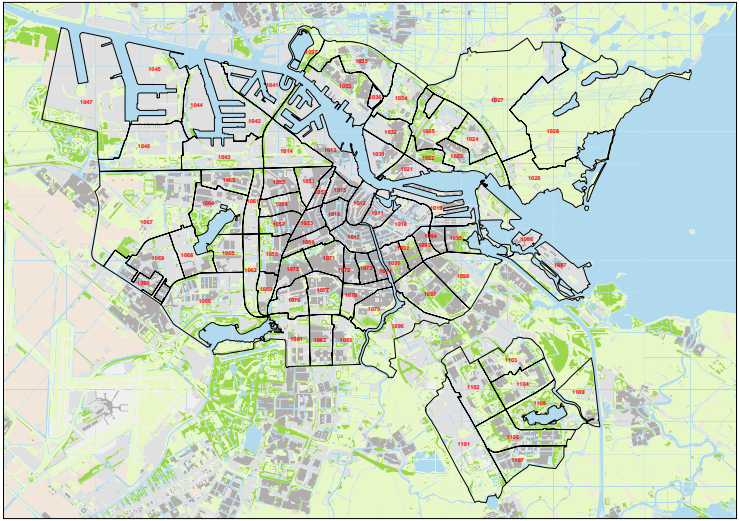

Figure S18: Map of Amsterdam postal code (PC4) areas.

## S2 Data

Data were obtained from Stichting HIV Monitoring, collected as part of the open ATHENA cohort of all patients in care in the Netherlands. The dataset includes includes the municipal health service (GGD) region of the patient at the time of registration to the cohort, or at their most recent registration update, based on their the postcode of their place of residency (PC4 code) either at time of registration to the cohort, or at their most recent registration update. PC4 is the most granular administrative city level in Amsterdam, with 12,000 residents on average per PC4 area and a number of residents ranging from 10 to 26,263. Figure S18 shows a map of the 81 Amsterdam PC4 areas. Amsterdam patients were identified as patients with a first or more recent registration in the Amsterdam GGD region.

The ATHENA database version was closed on March 31st 2019[2]. We obtained data for 19,204 patients from the Netherlands, with 7,773 of these having an Amsterdam postcode at first or last registration.

We leverage baseline data recorded at registration on year of birth, country of birth,

mode of transmission, date of death (if patient has died), date of AIDS diagnosis, date of ART start, date of last HIV negative test and date of first HIV positive test.

We also obtained datasets from the ATHENA cohort of partial HIV-1 polymerase (pol) sequences of Amsterdam patients, including date of sample, and of clinical data collected longitudinally of viral load measurements and CD4 counts.

In the study, we focus on infections estimated to have been acquired between 2014-2018 (see Section S3.1). We also consider MSM and heterosexual transmission groups only, since less than 2% of infections were in other transmission groups. Table S1 summarises patient characteristics for all Amsterdam individuals estimated to have been infected with HIV between 2014-2018, and those who have a viral sequence available. The cohort is predominantly male (92%), and MSM (86%). 41% of individuals were between 25-34 years old at their estimated time of infection. Less than 3% of individuals were estimated to have been infected aged 60 or older. 41% of individuals infected between 2014-2018 were born in the Netherlands, followed by 13% from South America and the Caribbean, which are predominantly individuals from Suriname and the Dutch Caribbean. Table S1 also reports characteristics of patients with a viral sequence available. Empirically comparing only those with a sequence with the complete Amsterdam cohort of all individuals infected between 2014-2018, indicates that those patients with a sequence are representative of the whole diagnosed population.

For each transmission group, we define each strata by place of birth, according to the main migrant populations in Amsterdam. For Amsterdam MSM and heterosexuals, respec-tively, these are,

$$\mathcal{M} = \{\text{Netherlands; W.Europe, North America and Oceania; Eastern and Central Europe;}$$
$$\text{Latin America and the Caribbean; Other}\}, \tag{S1a}$$

$$\mathcal{H} = \{\text{Netherlands; Sub-Saharan Africa; Latin America and the Caribbean; Other}\}. \tag{S1b}$$

Since we focus on infections acquired between 2014-2018, we define the study start and end time by,

$$\psi_{\text{start}} = 2014, \tag{S2a}$$

$$\psi_{\text{end}} = 2018. \tag{S2b}$$

## S3 Estimating HIV infection dates and undiagnosed infections

In this section, we first describe how we fit a model to clinical biomarker data to estimate the time from infection to diagnosis, and consequently the date of infection. Next, we describe how we fit a model to the posterior median estimates of the time to diagnosis, to estimate the proportion of Amsterdam infections which remained undiagnosed by the close of the study.

### S3.1 Estimating HIV infection dates

#### S3.1.1 Data

We define the complete cohort of patients registered in Amsterdam by $\mathcal{N}$. We first follow methods in [3] to estimate time from infection to diagnosis for individual $i \in \mathcal{N}$ by $w_i$. We use an indicator $R_i$ to denote transmission risk group of each individual, where,

$$R_i = \begin{cases} 1, & \text{if } i \text{ is Amsterdam MSM} \\ 0, & \text{if } i \text{ is Amsterdam heterosexual} \end{cases} \quad \text{(S3)}$$

We utilise clinical biomarker data for each patient on CD4 counts and viral loads, measured after diagnosis but before onset of AIDS or start of ART. As a caveat, we keep viral load measurements within one week of ART start, and CD4 counts within one month of ART start. This choice is supported by the fact that ART takes time to act. We denote CD4 counts by $y^c$, and viral loads by $y^r$, and encapsulate measurements for all $i$ individuals in a vector,

$$\mathbf{Y}_i^c = (y_{i1}^c, \ldots, y_{in_i^c}^c)^T \text{ and}$$
$$\mathbf{Y}_i^r = (y_{i1}^r, \ldots, y_{in_i^r}^r)^T. \quad \text{(S4)}$$

Each measurement is collected at an (unknown) time since infection,

$$\mathbf{t}_i^c = (t_{i1}^c, \ldots, t_{in_i^c}^c)^T \text{ and}$$
$$\mathbf{t}_i^r = (t_{i1}^r, \ldots, t_{in_i^r}^r)^T. \quad \text{(S5)}$$

We have clinical data prior to AIDS diagnosis or start of ART for 6,879 (88%) of patients. For the remaining 12% we are unable to estimate the time of infection. We then denote the time between diagnosis and each biomarker measurement by,

$$\mathbf{d}_i^c = (d_{i1}^c, \ldots, d_{in_i^c}^c)^T \text{ and}$$
$$\mathbf{d}_i^r = (d_{i1}^r, \ldots, d_{in_i^r}^r)^T. \quad \text{(S6)}$$

From this, we can then express the time from infection to measurement date in (S5) in terms of the estimated date of infection, $w_i$, and the time between diagnosis and each biomarker measurement as follows,

$$t_{ij}^c = d_{ij}^c + w_i \text{ and}$$
$$t_{ij}^r = d_{ij}^r + w_i. \quad \text{(S7)}$$

### S3.1.2 Model

We then use a bivariate linear mixed model for the joint distribution of the two biomarkers over time and denote their distribution by,

$$f(\mathbf{y}_i^c, \mathbf{y}_i^r | \mathbf{t}_i^c, \mathbf{t}_i^r), \tag{S8}$$

for the joint distribution of the two biomarkers over time. We place a uniform prior on $w_i$ over $(0, u_i)$, where $u_i$ is the interval between time at risk for each individual and HIV diagnosis. We take the risk onset date to be the maximum of the time between the last negative and test and diagnosis, and the time between the individual turning 15 years of age and diagnosis.

The posterior distribution of $w_i$ is as follows:

$$f(w_i | \mathbf{y}_i) = \frac{f(\mathbf{y}_i | w_i) f(w_i)}{\int_0^{u_i} f(\mathbf{y}_i | w_i) f(w_i) dw_i}, \tag{S9}$$

from which we estimate the median time from infection to diagnosis for $w_i$, and 95% credible intervals.

### S3.1.3 Estimated quantities

Then, if $T_i^{\text{diagnosis}}$ is the reported diagnosis date for individual $i$, we estimate their infection date, denoted by $T_i^{\text{infection}}$, with,

$$T_i^{\text{infection}} = T_i^{\text{diagnosis}} - w_i. \tag{S10}$$

Figure S1 shows the distribution of individual median estimates for time to diagnosis by the risk groups given by (S1a) and (S1b) for MSM and heterosexuals, respectively. Figure S2 plots the diagnosis date against the estimated infection date for all individuals diagnosed between 2014 and the May 2019. 95% credible intervals indicate uncertainty around indi-vidual level estimates from the model. We note that treatment guidelines changed in 2015 from starting ART based on CD4 count, which is measured every 6 months, to immediate ART initiation. Since we only consider biomarker measurements taken prior to ART start, as a result we have fewer biomarker measurements per individual for PLHIV diagnosed since 2015, which leads to larger uncertainty around date of infection.

## S3.2 Estimating the proportion of infections in 2014-2018 that were undiagnosed by May 2019

### S3.2.1 Data

We next sought to estimate the proportion of infections in 2014-2018 that remained undiagnosed by May 2019. The patient data is right-censored, so many recent infections may yet be undiagnosed in the patient data set. For this reason, we considered the subset of Amsterdam diagnoses that we estimated to have been acquired between 2010 and the end of 2012, since most infections acquired in this interval would have been diagnosed by early 2019 given typical disease progression [4]. We first define an indicator $U_i(\tau)$, which is a function of a given year $\tau$, in which,

$$U_i(\tau) = \begin{cases} 1, & \text{if } T_i^{\text{infection}} < \tau \\ 0, & \text{otherwise.} \end{cases} \tag{S11}$$

We then define the synthetic cohort of infections in 2010-2012 by S12.

$$\begin{aligned} \mathcal{C}^{MSM} &\subseteq \mathcal{N} : R_i = 1 \cap U_i(2010.0) = 0 \cap U_i(2013.0) = 1, \\ \mathcal{C}^{HSX} &\subseteq \mathcal{N} : R_i = 0 \cap U_i(2010.0) = 0 \cap U_i(2013.0) = 1. \end{aligned} \tag{S12}$$

We then consider individuals $k \in \mathcal{C}^{MSM}$ and $l \in \mathcal{C}^{HSX}$. For each transmission group, we defined each strata by place of birth given in equations S1a and S1b. Table S6 shows the characteristics of patients used to fit the model.

| Risk group | Place of birth | Amsterdam infections 2010-2012 | Median estimated time to diagnosis (years) [95% quantiles] |
|---|---|---|---|
| Amsterdam MSM | W.Europe, N.America, Oceania | 72 | 0.42 [0.05-3.41] |
| | E. & C. Europe | 31 | 0.88 [0.13-6.04] |
| | S. America & Caribbean | 81 | 1.04 [0.05-5.57] |
| | Netherlands | 295 | 0.56 [0.04-4.77] |
| | Other | 56 | 1.38 [0.12-4.97] |
| | **All** | **535** | **0.64 [0.04-4.97]** |
| Amsterdam heterosexual | Sub-Saharan Africa | 35 | 3.86 [0.33-6.8] |
| | S. America & Caribbean | 22 | 1.37 [0.14-5.68] |
| | Netherlands | 27 | 1.42 [0.07-6.16] |
| | Other | 13 | 1.6 [0.99-6.12] |
| | **All** | **97** | **2.22 [0.1-6.67]** |

Table S6: Patient characteristics for individuals with an estimated infection date between 2010-2012.

### S3.2.2 Hierarchical model

We fit a hierarchical Weibull model to the estimated times from infection to diagnosis ($w_i$) in Stan, for MSM and heterosexuals separately. For MSM, we denote the function $j(k)$ which takes as value the place of birth of individual $k$, as defined in equation (S1a). We estimate ethnicity-specific shape and scale parameters $\kappa_{j(k)\in\mathcal{M}}$ and $\lambda_{j(k)\in\mathcal{M}}$ which can borrow information from each other through a hierarchical prior distribution. For convenience when choosing priors, we re-parameterised the Weibull distribution in terms of its median and 80% quantile ($\log \chi^{50}_{j(k)}$, $\log \chi^{80}_{j(k)} - \log \chi^{50}_{j(k)}$). The quantile function for the Weibull distribution is given by Equation (S13).

$$Q(p; \kappa_{j(k)}, \lambda_{j(k)}) = \lambda_{j(k)}(-\log(1-p))^{1/\kappa_{j(k)}}. \tag{S13}$$

We then express the parameters of the Weibull distribution as follows:

$$\begin{aligned}
\kappa_{j(k)} &= \frac{\log(\log(5) - \log(2))}{\log \chi^{80}_{j(k)} - \log \chi^{50}_{j(k)}}, \\
\lambda_{j(k)} &= \exp\left(\log(\chi^{50}_{j(k)}) - \frac{\log(\log(2))}{\kappa_{j(k)}}\right),
\end{aligned} \tag{S14}$$

and then specify the Weibull model and its prior distribution as follows,

$$w_{j(k)} \sim \text{Weibull}\left(\frac{\log(\log(5) - \log(2))}{\log \chi^{80}_{j(k)} - \log \chi^{50}_{j(k)}}, \exp\left(\log(\chi^{50}_{j(k)}) - \frac{\log(\log(2))}{\kappa_{j(k)}}\right)\right), \quad \text{(S15a)}$$

$$\log(\chi^{50}_{j(k)}) \sim N(\mu_{\log \chi^{50}}, \sigma^2_{\log \chi^{50}}) \quad \text{(S15b)}$$

$$\log(\chi^{80}_{j(k)}) - \log(\chi^{50}_{j(k)}) \sim N(\mu_{\log \chi^{80} - \log \chi^{80}}, \sigma^2_{\log \chi^{80} - \log \chi^{50}}) \quad \text{(S15c)}$$

$$\mu_{\log \chi^{50}} \sim N(\log(Q(0.5)), 0.5) \quad \text{(S15d)}$$

$$\mu_{\log \chi^{80} - \log \chi^{50}} \sim N(\log(Q(0.5)) - \log(Q(0.8)), 0.5) \quad \text{(S15e)}$$

$$\sigma_{\log \chi^{50}} \sim \text{Exp}(2) \quad \text{(S15f)}$$

$$\log(\sigma_{\log \chi^{80} - \log \chi^{50}}) \sim N(0, 1), \quad \text{(S15g)}$$

where $Q(0.5)$ and $Q(0.8)$ are the empirical quantiles from the estimated times to diagnosis, for each transmission group.

The joint posterior distribution of model (S15a) was estimated in rstan with Stan version 2.19.3 using three Hamiltonian Monte Carlo chains with 2,000 samples each including a warmup of 500 samples. The models mixed well and had at most one divergence (Figures S19-S20).

### S3.2.3 Estimated quantities

We then estimate, for a given month $x \in \mathcal{X} = \{1, \ldots, 12\}$ (where $x = 1$ corresponds to January) and year $y \in \mathcal{Y} = \{2014, \ldots, 2018\}$ of infection, the probability of an MSM individual not being diagnosed by 1st May 2019, given their place of birth, as follows:

$$\delta^{(k)}_{j(k),x,y} = 1 - P\left(w_{j(k)} \leq \left(2019 + \frac{4}{12}\right) - \left(y + \frac{x}{12}\right) \mid \kappa_{j(k)}, \lambda_{j(k)}\right). \quad \text{(S16)}$$

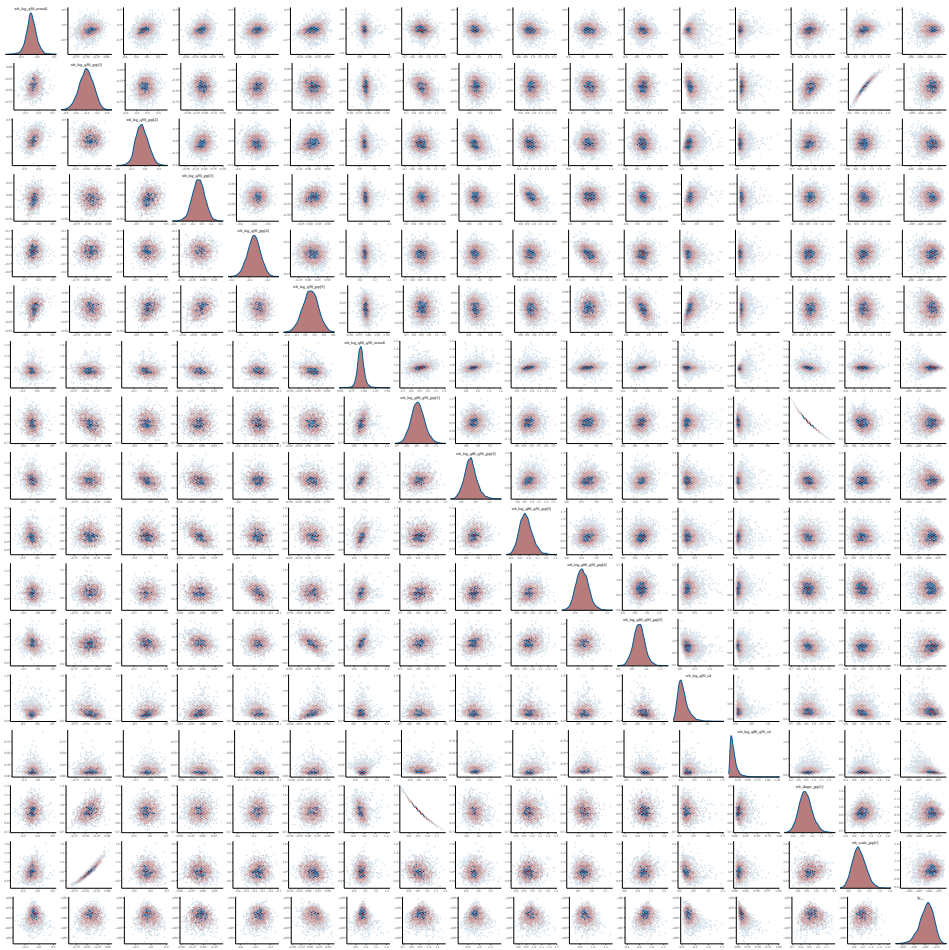

Figure S19: Pairs plot for MSM model.

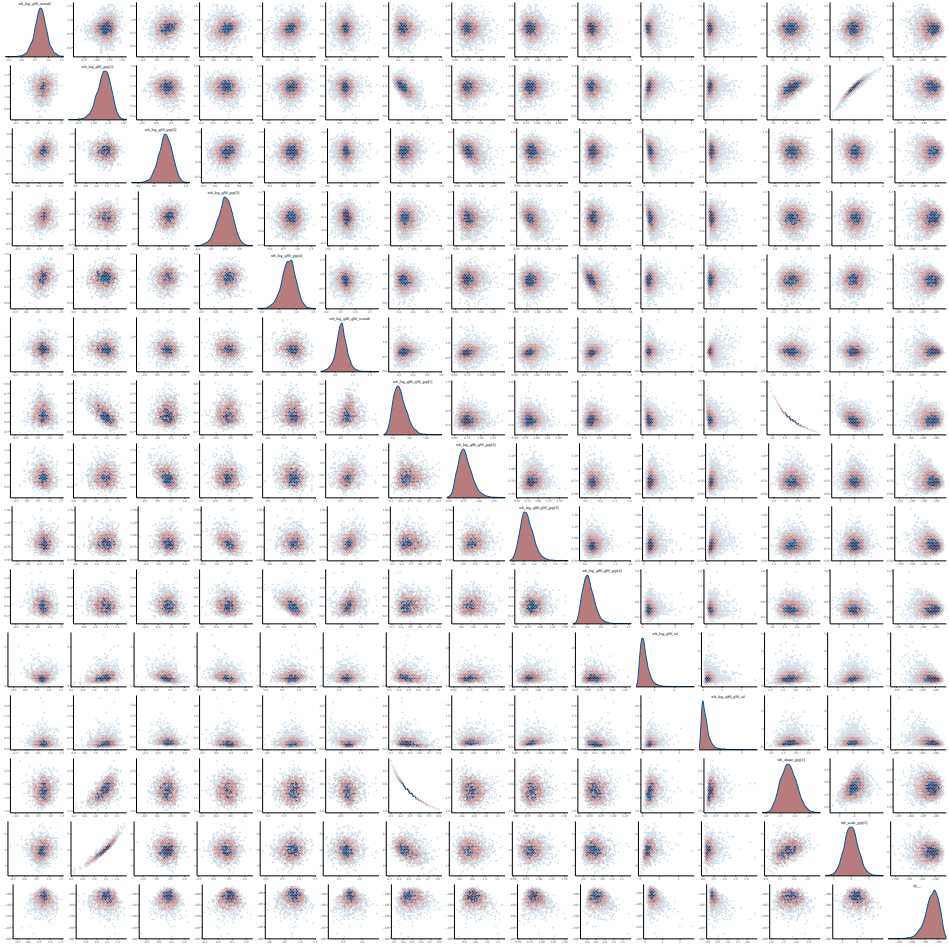

Figure S20: Pairs plot for heterosexual model.

We then calculate the average probability that an individual infected in each month of 2014-2018 remained undiagnosed by May 2019 with

$$\theta_{j(k)}^{(k)} = \frac{1}{12 * 5} \sum_{x \in \mathcal{X}} \sum_{y \in \mathcal{Y}} \delta_{j(k),x,y}^{(k)}. \tag{S17}$$

We then denote the number of diagnosed Amsterdam MSM infected in 2014-2018 and born in world region $M$ by $N_{Dm}$. Finally, we can estimate the total number of infections in 2014-2018 in Amsterdam MSM through,

$$N_I^{(k)} = \sum_{m \in M} \frac{N_{Dm}}{1 - \theta_m^{(k)}}, \tag{S18}$$

and obtain numerical estimates of $N_I^{(k)}$ via the Monte Carlo samples from the joint posterior and the calculated proportions $\theta_m$ of undiagnosed infections. Poster median estimates and 95% credible intervals of (S16)-(S18) are obtained by summarising the set of Monte Carlo samples after the transformations. The model for heterosexuals is formulated analogously.

Figure S21 shows the estimated Weibull distributions for the time to diagnoses, stratified by MSM and heterosexuals and place of birth. The empirical cumulative distribution functions (CDFs) of the times to diagnoses are for comparison shown as step functions (black). The fits were good, with the empirical CDFs generally lying within the 95% posterior intervals of the fitted CDFs for all risk groups. Figure S3 summarises the total number of estimated infections acquired between 2014-2018, the subset of those that were diagnosed by May 2019, and the subset of those which have a viral sequence available. The sequence sampling fraction is shown above each bar.

### S3.3 Sensitivity analysis: using only data on last negative tests

We carried out several sensitivity analyses to explore the impact of alternative approaches to estimating infection dates on the proportion of Amsterdam infections in 2014-2019 that are estimated to have remained undiagnosed by 2019. We first considered estimating the date of infection as the midpoint between last negative HIV test and first positive HIV test, where available. We therefore only considered patients with a last negative HIV test to fit the model for the time to diagnosis distributions. In contrast, the approach taken to estimating the infection date in the main analysis considers the time at risk to be either the time since last negative HIV test, or the time since the patient was 15 years old where a last negative test is not available. Based on the midpoint estimates, each individual was classified to have been infected before or after 2014 in analogy to equation (S11). We had 266 patients across the synthetic cohorts defined by equation (S12), compared with 632 when using the estimated date of infection. This is reflective of the fact many individuals do not have a reported last negative test date.

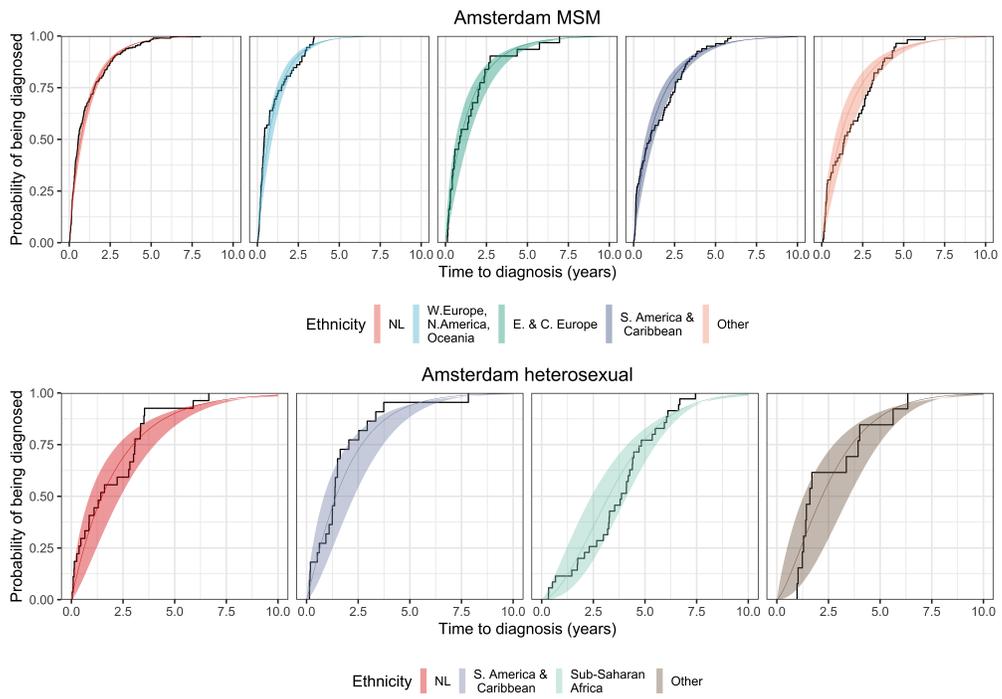

Figure S21: Posterior median cumulative distribution functions (CDFs) (line in colours) and 95% credible intervals (ribbon in colours) are shown along with the empirical CDF (steps in black).

Figure S22 compares the estimated proportion of undiagnosed Amsterdam infections obtained as in the main analysis from all biomarker data from all individuals (Figure S22a) to that obtained when using only midpoint estimates from seroconverters (Figure S22b). Estimates are compared by year of infection for each risk group. When using data only from the seroconverters, the estimated proportions of undiagnosed individuals are much smaller. This is likely driven by the fact we excluded patients without a last negative test, who may have typically had longer estimated times to diagnosis. This was also observed by Ratmann et al.[5]. There are also considerably fewer data points, particularly among heterosexuals, resulting in elevated uncertainty in these estimates. Figure S23 shows our estimates for the total number of infected individuals in Amsterdam. Clearly, whilst the estimates are more conservative where we use midpoint estimates than we find using the estimated times to diagnosis (see Figure S3), we still find a substantial proportion of individuals to be undiagnosed by May 2019.

## S3.4 Sensitivity analysis: different classification rules

Second, we considered alternative classification rules for grouping individuals into infected before or after 2014. Specifically, we considered as cutoff point in equation (S10) the 30% (and 40%) quantile of the posterior distributions of individual time to diagnosis times, in place of the posterior median estimate. The model for estimating times to diagnosis from biomarker data has been previously validated with good accuracy on classifying individuals into two groups of those infected before or after a certain cut-off point [3]. Here, we nonetheless performed these sensitivity analyses to quantify the potential impact of hypothetical, systematic bias in the time to diagnosis estimates on the estimated proportion of undiagnosed Amsterdam infections.

Figures S22c and S22d shows the estimates for the undiagnosed by year. As expected, the 30% and 40% quantiles result in shorter times to diagnosis, and therefore we estimate a smaller proportion of individuals to remain undiagnosed by May 2019. Specifically, we find 16% [15%-18%] of all infections between 2014-2018 undiagnosed by May 2019 using the 30% quantiles, and 21% [19%-23%] of infections using the 40% quantiles. Figures S23b and S23c show the estimated total infected population.

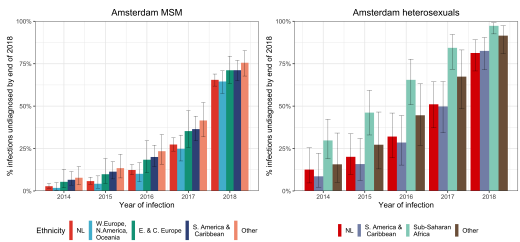
(a) Using all biomarker data from all individuals.

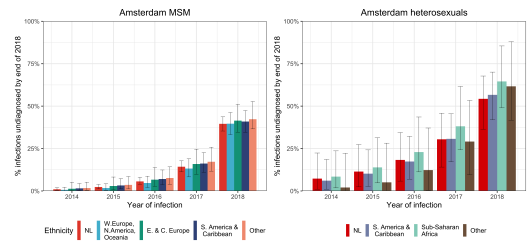
(b) Using midpoint estimates from seroconverters.

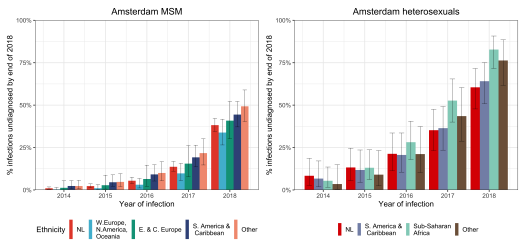
(c) Using 30% quantile of posterior distribution of time to diagnosis.

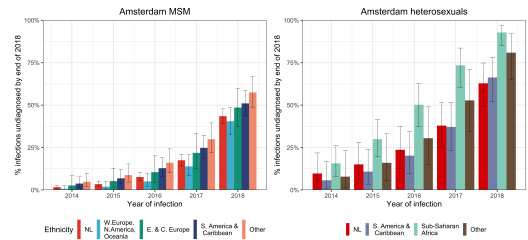
(d) Using 40% quantile of posterior distribution of time to diagnosis.

Figure S22: Estimated proportion of Amsterdam infections in 2014-2018 which remained undiagnosed by May 2019, by year of infection.

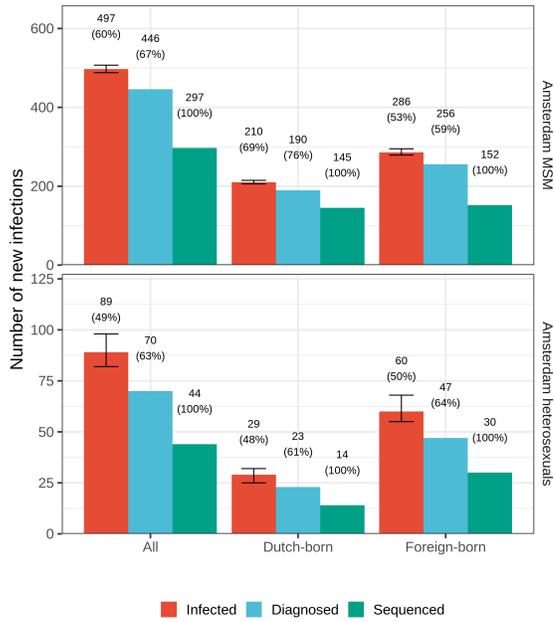

(a) Using midpoint estimates from seroconverters.

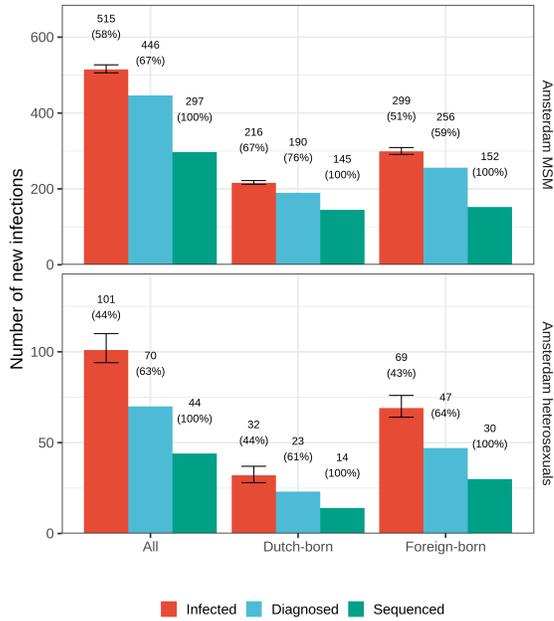

(b) Using 30% quantile of posterior distribution of time to diagnosis.

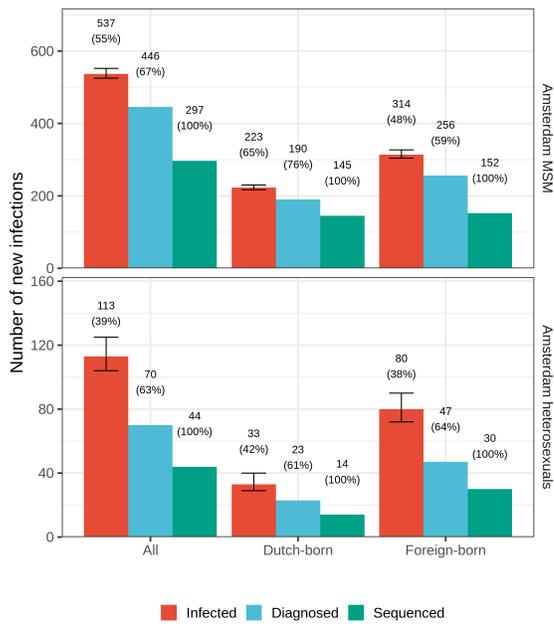

(c) Using 40% quantile of posterior distribution of time to diagnosis.

Figure S23: Estimated Amsterdam infections in 2014-2018. Estimates of the total number of individuals resident in Amsterdam that were infected in 2014-2018 are shown along with the subset of individuals that were diagnosed, and the subset of those for who at least one viral sequence is available. Posterior median estimates (bars, and number on top of bar) are shown along with 95% credible intervals. The posterior median proportion of individuals with a viral sequence is also shown (proportion on top of bar).

## S3.5 Sensitivity analysis: Estimates from ECDC modelling tool

Third, we considered utilising estimates for Amsterdam from the ECDC HIV modelling tool [6]. We used model estimates of the estimated infections per year among MSM and heterosexuals, respectively, and the estimated proportion of infections diagnosed within $1, \ldots, 15$ years. Estimates were only available by transmission group, but not place of birth. Estimates also also only available by year, so we estimate the proportion of infections acquired between 2014-2018 undiagnosed by the end of 2019.

If $\Delta_y$ are the estimated number of individuals infected in year $y$, and $\delta_{y,z}$ is the probability of an individual infected in year $y$ being diagnosed in year $z \in \{y, \cdots, 2018\}$. Then, the proportion of individuals infected between 2014-2018 who are undiagnosed by the end of 2019, are estimated by:

$$\theta = \sum_{y \in \mathcal{Y}} \gamma_i (1 - \sum_{z=y}^{2018} \delta_{y,z}) \tag{S19}$$

where $\gamma_i$ are weights according to the proportion of individuals infected in year $y$:

$$\gamma_y = \frac{I_y}{\sum_{x \in \mathcal{Y}} I_x} \tag{S20}$$

Estimates of undiagnosed were similar for MSM (28.7%) of infections were undiagnosed by May 2019, but considerably higher among heterosexuals compared to the estimates from the Weibull model (62.2%).

## S4 Viral phylogenetic analyses

### S4.1 Multiple sequence alignment

We used partial HIV pol sequences from Amsterdam and the rest of the Netherlands from the ATHENA cohort and aligned these to the reference genome HXB2 [7] using `Virulign` [8].

Sequences which failed to align were aligned globally using Mafft version 7 [9]. Nucleotide positions which were missing for most sequences, or not in the reference sequence HXB2 were removed. Known antiretroviral resistant mutations were masked using the R package `big.phylo`[10]. The final alignment was 1302nt in length. We carried out some manual curation of the alignment, removing all gaps and resolving sequences which did not align correctly. We then classified sequences by subtype using `COMET` [11] and verified any which were uncertain with `REGA` v3.0 [12].

We downloaded 82,708 background sequences from the Los Alamos HIV-1 sequence database on 27th February 2020, specifying fragments in the POL region longer than 1300nt. We then used the Basic Local Alignment Search tool (`BLAST`, https://blast.ncbi.nlm.nih.gov/Blast.cgi) to identify the top 20 closest background sequences to each of the Dutch sequences, which we kept and aligned to the Dutch sequences using the HXB2 reference sequence. We created alignments by subtype, excluding the least common subtypes with fewer than 50 Dutch and background sequences.

## S4.2 Reconstruction of city transmission chains

We used `FastTree` v2.1.8 to reconstruct phylogenetic trees for each subtype [13]. We then assigned labels to each sequence. Sequences from Amsterdam were labelled according to their risk group, sequences from the rest of the Netherlands (excluding Amsterdam) were labelled as such, and background sequences were labelled according to the country they originated from. The geographic regions for the MSM trees were,

$$\mathcal{N} = \{\text{Amsterdam MSM}, \text{Amsterdam non-MSM}, \text{Netherlands}, \text{Africa}, \text{Western Europe},$$
$$\text{Eastern Europe and Central Asia}, \text{North America}, \text{Latin America and the Caribbean},$$
$$\text{Dutch Caribbean and Suriname}, \text{Middle East and North Africa}, \text{Asia and Oceania}\},$$
(S21)

and similarly for heterosexual trees,

$$\begin{aligned}\mathcal{O} = \{&\text{Amsterdam heterosexual}, \text{Amsterdam non-heterosexual}, \text{Netherlands}, \text{Africa},\\ &\text{Western Europe}, \text{Eastern Europe and Central Asia}, \text{North America},\\ &\text{Latin America and the Caribbean}, \text{Dutch Caribbean and Suriname},\\ &\text{Middle East and North Africa}, \text{Asia and Oceania}\},\end{aligned} \quad \text{(S22)}$$

We then used `phyloscanner` v1.8.0 [14] to assign one of the state labels to each viral lineage in the reconstructed phylogenies. Figures S5-S15 show the annotated phylogenetic trees for all major subtypes and circulating recombinant forms that circulate in Amsterdam. From the annotated trees, we extracted the viral phylogenetic subgraphs that were assigned to Amsterdam individuals. We assume that viral phylogenetics correctly assigns individuals into subgraphs, which we interpret as the observed parts of distinct city-level transmission chains.

## S5 Branching process model of partially observed, growing transmission chains

In this section, we describe how we build on existing methods to model the growth of the existing and newly introduced transmission chains. Utilising the phylogenetic subgraph data described in Section S4.2, we show how we can model their growth from a point in time, rather than from the first introduction, by utilising the number of infectious cases in the subgraph at a given point in time, and how many new cases were generated from those infectious cases. We also describe the model likelihood of new transmission chains which emerged.

We model the growth of transmission chains using putative infection dates, estimated in S3.1. For individuals with no estimate for date of infection, due to missing clinical biomarker data after diagnosis, we subtracted the posterior median time to diagnosis for an individual estimated using the model described in equation (S15a) in the corresponding migrant group, defined by equations (S1a)-(S1b).

## S5.1 Probability that $m$ index cases collectively generate $i$ new infections

We model the spread of HIV transmission chains that are characterised by reproduction numbers below one, through branching processes characterised by Negative Binomial offspring distributions [15]. A central component of branching process theory is the probability generating function $Q(s) = \sum_{i=0}^{\infty} q_i s^i$, where $q_i$ is the probability that one individual generates $i$ new infections in one generation, and $q_0$ is the probability that one individual generates no further infections. For our purposes, we will use two fundamental formulae. First, the $k$th derivative of $Q$ is

$$\frac{d}{d^k s} Q(s) = \sum_{i=k}^{\infty} \frac{i!}{(i-k)!} \, q_i \, s^{i-k}, \tag{S23}$$

and so the probability $q_k$ is recovered through

$$q_k = \frac{1}{k!} \frac{d}{d^k s} Q(0). \tag{S24}$$

Second, the $k$th coefficient of $Q^2(s)$ is

$$\sum_{j=0}^{k} q_j q_{k-j}, \tag{S25}$$

which is the probability that two individuals collectively generate $k$ new infections. Thus, the probability that $m$ index cases collectively generate $i$ new infections is given by the $i$th coefficient of $Q^m(s)$. We denote this probability by

$$p(i|m) = \frac{1}{i!}\frac{d}{d^i s}Q^m(0). \tag{S26}$$

We consider a Negative Binomial offspring distribution, parameterised in terms of the mean $\mu$ and dispersion parameter $\phi$, so that its variance is given by $\mu(1 + \mu/\phi)$. Thus, as $\phi$ tends to zero, $\mu/\phi$ increases, and so does the variance to mean ratio $(1 + \mu/\phi)$. This means that the Negative Binomial can simultaneously model average reproduction numbers as well as additional heterogeneity in the number of new infections per generation, that goes beyond the variation described by a Poisson offspring distribution. The probability generating function of the Negative Binomial offspring distribution is

$$Q(s) = \left(1 + \frac{\mu}{\phi}\right)^{-\phi}. \tag{S27}$$

Thus, we have that the probability that $m$ index cases generate $i$ new infections is

$$p(i|m,\mu,\phi) = \frac{1}{i!}\frac{d}{d^i s}Q^m(0) \tag{S28a}$$

$$= \frac{1}{i!}\left(\prod_{k=0}^{i-1}(\phi m + k)\left(\frac{\mu}{\phi}\right)^i\left(1 + \frac{\mu}{\phi}\right)^{-\phi m - i}\right) \tag{S28b}$$

$$= \frac{(\phi m + i - 1)!}{i!\,(\phi m - 1)!}\left(\frac{\mu}{\phi}\right)^i\left(1 + \frac{\mu}{\phi}\right)^{-\phi m - i} \tag{S28c}$$

$$= \frac{(\phi m + i - 1)!}{i!\,(\phi m - 1)!}\left(\frac{\phi}{\mu + \phi}\right)^{\phi m}\left(1 - \frac{\phi}{\mu + \phi}\right)^i, \tag{S28d}$$

where $m = 1, 2, \ldots$ are fixed, and the number of new infections takes on values $i = 0, 1, \ldots$. It is helpful to note that equation (S28a)-(S28d) has an intuitive interpretation, it is a Negative Binomial with mean $\mu m$ and dispersion parameter $\phi m$, which we denote by

$$p(i|m,\mu,\phi) = \text{NegBin}(i \mid \mu m, \phi m), \tag{S29}$$

where $m = 1, 2, \ldots$ are fixed, and the number of new infections takes on values $i = 0, 1, \ldots$.

Equivalently, we can express the probability that $m$ index cases result in a total number of $n$ cases through

$$\tilde{p}(n|m,\mu,\phi) = \frac{(\phi m + n - m - 1)!}{(n-m)!\,(\phi m - 1)!} \Big(\frac{\phi}{\mu+\phi}\Big)^{\phi m} \Big(1 - \frac{\phi}{\mu+\phi}\Big)^{n-m}, \tag{S30}$$

or more simply

$$\tilde{p}(n|m,\mu,\phi) = \mathrm{NegBin}(n - m \mid \mu m, \phi m), \tag{S31}$$

where $m = 1, 2, \ldots$ are fixed, and the number of total cases are $n = m, m+1, \ldots$.

## S5.2 Probability that $m$ index cases result in a transmission chain with $i$ new infections

Transmission chains require that infections occur in a particular order, while in contrast Equations ((S28a)-(S28d)) do not impose in what generation how many infections occur. For example, with one index case $m = 1$ and a total size $n$, Equation (S30) quantifies the probability that $n - 1$ new infections occur, but there is no constraint that the index case generates at least one new infection in the next generation.

Dwass [16] derived the correction factor, and the probability that a transmission chain with $m$ index cases has $i$ new infections, or equivalently $n$ cases, is

$$c(i|m,\mu,\phi) = \frac{m}{m+i}\, p(i|m,\mu,\phi) \tag{S32a}$$

$$\tilde{c}(n|m,\mu,\phi) = \frac{m}{n}\, p(n-m|m,\mu,\phi), \tag{S32b}$$

where $m = 1, 2, \ldots$, $i = 0, 1, \ldots$, and $n = m, m+1, \ldots$.

## S5.3 Probability that $m$ index cases result in subgraphs with $i$ sampled, new infections

In practice, only a subset of new infections are captured in viral phylogenies because only a subset of infections are diagnosed, and of those only a subset have virus sequenced. We make two assumptions. First, infections are missing independently of each other with the same probability $1-\rho$, so $\rho$ is the sampling probability of infections. Second, uncertainty in $\rho$ can be quantified within several percentage points through surveillance data and/or modelling; we use this assumption later to ensure that the remaining parameters are statistically identifiable.

Then, the probability of observing $i$ individuals in a subgraph that has $m$ known index cases is

$$\begin{aligned} c_{\text{obs}}(i|m,\mu,\phi,\rho) &= \sum_{k=i}^{\infty} \Big( \text{Bin}(i|k,\rho) c(k|m,\mu,\phi) \Big) \\ &= \sum_{k=i}^{\infty} \Big( \text{Bin}(i|k,\rho) \frac{m}{m+k} \text{NegBin}(k|\mu m, \phi m) \Big), \end{aligned} \tag{S33}$$

where $m = 1, 2, \ldots$, $i = 0, 1, \ldots$, and $c(k|m,\mu,\phi)$ is from Equation (S32a). It is possible that an observed subgraph has $m$ index cases by a particular study start time $\psi_{\text{start}}$ and no new infections between $\psi_{\text{start}}$ and $\psi_{\text{end}}$, as defined in equations (S2a)-(S2b)), and the probability of observing one such subgraph is $c_{\text{obs}}(0|m,\mu,\phi,\rho)$.

## S5.4 Probability that emergent subgraphs have $n$ sampled cases

Some observed subgraphs are emergent in the sense that they consist of individuals that were all diagnosed after the study start time $T$. In this case, Equation (S33) cannot be used because it assumes that subgraphs contain at least one index case prior to the study start time $T$. We assume that emergent subgraphs are seeded by one index case, which for example ignores the possibility that sexual partners infected each other and then moved to Amsterdam, and seeded a new transmission chain in Amsterdam. The probability of observing an emergent transmission chain of size $n$ is given by

$$\begin{aligned}
\tilde{c}_{\text{obs}}(n|m=1,\mu,\phi,\rho) &= \frac{\sum_{z=n}^{\infty}\left(\text{Bin}(n|z,\rho)\tilde{c}(z|m=1,\mu,\phi)\right)}{1-\sum_{z=1}^{\infty}\left(\text{Bin}(0|z,\rho)\tilde{c}(z|m=1,\mu,\phi)\right)} \\
&= \frac{\sum_{z=n}^{\infty}\left(\text{Bin}(n|z,\rho)\tilde{c}(z|m=1,\mu,\phi)\right)}{1-\sum_{z=1}^{\infty}\left((1-\rho)^z \tilde{c}(z|m=1,\mu,\phi)\right)} \\
&= \frac{\sum_{z=n}^{\infty}\left(\text{Bin}(n|z,\rho)\frac{1}{z}\text{NegBin}(z-1|\mu,\phi)\right)}{1-\sum_{z=1}^{\infty}\left((1-\rho)^z \frac{1}{z}\text{NegBin}(z-1|\mu,\phi)\right)},
\end{aligned} \quad \text{(S34)}$$

where unlike Equation (S33), $n = 1, 2, \ldots$ may include in the count the index case (if it is sampled), and $\tilde{c}(z|m=1,\mu,\phi)$ is from Equation (S32b). The denominator corrects for the event that the index case and all new infections in an emergent chain are unsampled, which is possible with non-zero probability, but always unobserved.

## S5.5 Likelihood of the growth distribution of phylogenetic subgraphs

We now describe the likelihood of the growth distribution of viral phylogenetic subgraphs, which throughout we identify as the observed parts of distinct city-level transmission chains. In what follows, for brevity, we only consider one transmission group and omit reference to this transmission group. All equations are analogous for the other transmission group.

We start with the viral phylogenetic subgraphs in the viral phylogeny of one subtype, and omit for brevity also any indication of that subtype. The data consist of a two-dimensional array $\mathbf{x}$, where $x_{mi}$ denotes the number of subgraphs that had $m$ index cases at the study start time $\psi_{\text{start}}$ and $i$ sampled, new infections by the study end time $\psi_{\text{end}}$. Here, $m = 1, \ldots, M$ and $i = 0, \ldots, I$ where $M$ denotes the largest number of index cases observed, and $I$ denotes the largest number of new infections observed. In addition, the data consist of a one-dimensional array $\tilde{\mathbf{x}}$, where $\tilde{x}_n$ denotes the number of emergent subgraphs that have $n$ sampled cases during the study period. Here, $n = 1, ..., N$, because at least one case needs to be sampled in order to observe the corresponding subgraph.

Then, we associate the following log-likelihood to the growth distributions of pre-existing and emergent subgraphs,

$$l(\mathbf{x}, \tilde{\mathbf{x}} | \mu, \phi, \rho) = \bigg( \sum_{m=1}^{M} \sum_{i=0}^{I} x_{mi} \log c_{obs}(i | m, \mu, \phi, \rho) \bigg) + \bigg( \sum_{n=1}^{N} \tilde{x}_n \log \tilde{c}_{\text{obs}}(n | m = 1, \mu, \phi, \rho) \bigg). \tag{S35}$$

The log-likelihood thus involves infinite sums through equations (S33) and (S34). We approximated these by summing up to the $10 * I * (N_D/N_S)$th term, where $N_D$ are the number of diagnosed individuals and $N_S$ are the number of sequenced individuals, so $I * (N_D/N_S)$ is the expected number of individuals in the transmission chain that corresponds to the largest observed subgraph. In applying this log-likelihood, we assume that (1) all transmission chains have reached their final size by the end of the study period, i.e. that they are complete; (2) that all emergent transmission chains have one index case; (3) that each case has an equal and independent probability of being sampled.

Next we consider the joint likelihood that arises from consideration of viral subgraphs of the same transmission group (e.g. MSM or heterosexual individuals) across all HIV subtypes or circulating recombinant forms. Since the number of subgraphs and new cases acquired between 2014-2018 are very small for some subtypes, we aggregate the subgraph size distributions for non-B subtypes. We index subtypes B and non-B by $s = 1, \ldots, S$, where $S = 2$, and denote the corresponding subgraph growth distributions by $\mathbf{x}_s$, and $\tilde{\mathbf{x}}_s$.

Then, we model the log-likelihood of all the data for one transmission group through

$$
\begin{aligned}
ll &= \sum_{s=1}^{S} l(\mathbf{x}_s, \tilde{\mathbf{x}}_s | \mu_s, \phi_s, \rho_s) \\
&= \sum_{s=1}^{S} \Bigg[ \Bigg( \sum_{m=1}^{M} \sum_{i=0}^{I} x_{mi} \log c_{obs}(i|m, \mu_s, \phi_s, \rho_s) \Bigg) + \\
&\quad \Bigg( \sum_{n=1}^{N} \tilde{x}_n \log \tilde{c}_{\text{obs}}(n|m=1, \mu_s, \phi_s, \rho_s) \Bigg) \Bigg],
\end{aligned}
\qquad (S36)
$$

where the $\mu_s$, $\phi_s$, $\rho_s$ are specific to the corresponding transmission group and subtype.

## S6 Bayesian inference

We estimate city-level transmission dynamics, the growth distribution of transmission chains, and the proportion of locally acquired infections through the log-likelihood (S36) of phylogenetically observed subgraphs.

### S6.1 Preliminaries

#### S6.1.1 Number of index cases

For each individual $i$ in the cohort $\mathcal{N}$, if $r_i$ is their last viral load measurement taken before 2014, we define them to be not virally suppressed by 2014 through,

$$
S_i = \begin{cases} 1, & \text{if } T_i^{\text{infection}} < 2014 \cap r_i > 100 \\ 0, & \text{otherwise.} \end{cases}
\qquad (S37)
$$

Then, for each observed subgraph $j$ where $(j = 1, ..., A)$, $m_j$ are the observed index cases, we count the number of individuals infected by, but who were not virally suppressed, by the start of 2014. For example for MSM, if $\mathcal{C}^{MSM}$ is the subset of MSM in Amsterdam,

$$
\mathcal{C}^{MSM} \subseteq \mathcal{N} : R_i = 1,
\qquad (S38)
$$

the number of observed index cases in subgraph $j$ is,

$$m_j = \sum_{k \in \mathcal{C}^{MSM}} S_{jk}, \tag{S39}$$

and $m_j > 0$. We count analogously for heterosexuals. The actual number of index cases $m_j^* \sim \text{NegBinom}(m_j, \nu)$, where $\nu$ is the sampling fraction of individuals who were not virally suppressed by 2014. We estimate the true number of index cases under complete sampling $m_j^*$ by,

$$E(m_j^*) = \frac{m_j}{\nu}, \quad i = 1, ..., A \tag{S40}$$

When $m_j = 0$, estimate $m_j^*$ from the mode of the pmf for the distribution $\text{Binomial}(0; m_j^*, \nu)$.

### S6.1.2 Number of subgraphs with no new infections

For the subgraphs in which no individuals were not virally suppressed by 2014 (i.e. no observed index case), and no observed new case between 2014-2018, were not included in the subgraph sizes and assumed to have died out.

### S6.2 Hierarchical model

The parameters of the model (S36) are the subtype-specific mean parameters of the offspring distributions, $\mu_1, \ldots, \mu_S$, the dispersion parameters $\phi_s$ and the sampling parameter $\rho$. To estimate the $\mu_1, \ldots, \mu_S$, we borrow information across subtypes through a hierarchical prior distribution. We interpret the mean parameters of the offspring distributions as the effective reproduction numbers during the study period for the corresponding subtype. The variance-to-mean ratio of the Negative Binomial offspring distribution is $1 + \mu_s/\phi_s$ and measures the degree of dispersion of the size distribution of the transmission chains. For ease of inference, we re-parameterize the dispersion parameter into the variance-to-mean ratio minus one and also specify a hierarchical prior distribution,

$$v_s = \mu_s/\phi_s. \tag{S41}$$

The log posterior density is given by

$$\begin{aligned}
&\log p\Big(\mu^s, v, \rho | \mathbf{x}_s, \tilde{\mathbf{x}}_s, s = 1, \ldots, S\Big) \\
&\propto \sum_{s=1}^{S} ll\Big(\mathbf{x}_s, \tilde{\mathbf{x}}_s | \mu_s, \mu_s v, \rho\Big) + \sum_{s=1}^{S} \log p(\mu^s) + \log p(v) + \log p(\rho)
\end{aligned} \tag{S42}$$

where the prior densities are specified as follows. For the effective reproduction numbers, we specified the normal-normal two-level prior

$$\log \mu_s \sim N(\log \mu, \sigma^2)$$
$$\log \mu \sim N(\hat{\mu}_{\log \text{MLE}}, 0.3) \tag{S43}$$
$$\sigma \sim \text{Exp}(0.1).$$

The hyperprior of the grand mean was centred on the maximum likelihood estimate $\log \hat{\mu}_{\text{MLE}} = \log(1 - 1/\bar{x})$, where $\bar{x}$ is the average subgraph size [17]. The hyperprior of the grand standard deviation $\sigma$ was specified by considering the differences in the log maximum likelihood estimates $\log \hat{\mu}_{\text{MLE}}$ for each subtype.

For the variance-to-mean ratio, we specified

$$v_s \sim \text{Exp}(v), v \sim \text{Exp}(1), \tag{S44}$$

where 1 is the rate parameter for the exponential distribution. For the sampling parameter, we specified

$$\rho \sim \text{Beta}(N_S + 0.5, (N_D/(1-\theta) - N_S) + 0.5), \tag{S45}$$

where $N_S$ are the number of sequenced individuals, $N_D$ are the number diagnosed and $\theta$ are the proportion of undiagnosed individuals.

### S6.3 Numerical inference

The joint posterior distribution was estimated using Stan version 2.19.3 across three chains, each with 2,000 samples.

The models mixed well; Figures S26-S27 shows the pairs plot of parameters for the MSM and heterosexual models, respectively. We note that we did not observe multiplicative non-identifiabilities (banana shape) between the reproduction rate R0 and the variance-to-mean ratio, as found by Blumberg and Lloyd-Smith [17].

## S7 Target quantities derived from fitted model

### S7.1 Estimated number of new cases in transmission chains since 2014

To estimate the actual number of new infections in transmission chains since 2014 from the phylogenetically observed subgraphs, we use the model fits in combination with the size equations (S32a) and (S32b) to obtain the posterior predictive number of new cases in a transmission chain with $m = 1, \ldots$ index cases in 2014. For emergent chains, we assume as before that there was one index case since 2014. Specifically, we have

$$p(i^* | \mathbf{x}, \tilde{\mathbf{x}}, m) = \int c(i^* | m, \mu, \phi) p(\mu, \phi | \mathbf{x}, \tilde{\mathbf{x}}) d(\mu, \phi) \tag{S46}$$

where $i^* = 0, 1, \ldots$, and for ease of notation we have dropped the suffixes for different subtypes, transmission groups, or time intervals. We approximate equation (S46) numerically from $k = 1, \ldots, K$ Monte Carlo samples $\mu^{(k)}, \phi^{(k)}$ of the joint posterior distribution by generating samples from

$$i^{*(k)} \sim c(i^* | m, \mu^{(k)}, \phi^{(k)}), \quad k = 1, \ldots, K. \tag{S47}$$

This is easily done since the inference algorithm already tabulates the probabilities $c(i^*|m, \mu^{(k)}, \phi^{(k)})$ for $i^* = 0, 1, \ldots$.

Equation (S47) allows us to generate one Monte Carlo sample of the actual growth of all transmission chains. We denote the number of all pre-existing phylogenetically observed transmission chains with at least one index case by

$$|\mathbf{x}| = \sum_{m=1}^{M} \sum_{i=1}^{I} x_{mi}, \tag{S48}$$

and index each of them through $j^x = 1, \ldots, |\mathbf{x}|$. Correspondingly we denote the number of emergent subgraphs by

$$|\tilde{\mathbf{x}}| = \sum_{n=1}^{N} \tilde{x}_n. \tag{S49}$$

A certain proportion of emergent transmission chains remains phylogenetically unobserved owing to incomplete sampling. In our model, the probability that an emergent transmission chain is entirely unobserved is given by

$$\rho_{\text{not-obs}}^{e} = \sum_{z=1}^{\infty} (1-\rho)^z \frac{1}{z} c(z-1|m=1, \mu, \phi), \tag{S50}$$

as in Equation (S34). Thus, the expected number of emergent transmission chains is $|\tilde{\mathbf{x}}|/(1 - \rho_{\text{not-obs}}^{e})$. We obtain a Monte Carlo estimate of (S50) by plugging in our estimates of the joint posterior density,

$$\rho_{\text{not-obs}}^{e(k)} = \sum_{z=1}^{\infty} (1-\rho^{(k)})^z \frac{1}{z} c(z-1|m=1, \mu^{(k)}, \phi^{(k)}). \tag{S51}$$

Using Equation (S51), we can predict the number of completely unobserved, emergent subgraphs through

$$N_{\text{not-obs}}^{*(k)} \sim \text{NegBin}\Big(|\tilde{\mathbf{x}}|, \rho_{\text{not-obs}}^{e(k)}\Big), \tag{S52}$$

where the Negative Binomial is specified in terms of the number of failures and success probabilities, with mean $\big(|\tilde{\mathbf{x}}|(1-\rho_{\text{not-obs}}^{e(k)})\big)/\big(1-\rho_{\text{not-obs}}^{e(k)}\big)$, so that the mean of $|\tilde{\mathbf{x}}| + N_{\text{not-obs}}^{*(k)}$ equals as desired $|\tilde{\mathbf{x}}|/(1-\rho_{\text{not-obs}}^{e(k)})$. We index all emergent transmission chains through $j^e = 1, \ldots, |\tilde{\mathbf{x}}| + N_{\text{not-obs}}^{*(k)}$, and note that the number of emergent transmission chains is uncertain.

Then, the actual number of new cases since 2014 in the chain corresponding to the observed, pre-existing subgraph $j^x$ is predicted by sampling $i_{j^x}^{*(k)} \sim c(\cdot | m_{j^x}, \mu^{(k)}, \phi^{(k)})$, where $m_{j^x}$ is the number of index cases in the corresponding subgraph. Similarly, the actual number of new cases since 2014 of the chain corresponding to the emerging transmission chain $j^e$ is predicted by sampling $i_{j^e}^{*(k)} \sim \tilde{c}(\cdot | 1, \mu^{(k)}, \phi^{(k)})$, and then calculating $n_{j^e}^{*(k)} = i_{j^e}^{*(k)} + 1$. For the emergent subgraphs, we add 1 since we assume as before that the index case occurred after 2014. Aggregating these sizes, we predict the size distribution of the number of chains with $i$ new cases since 2014 by

$$x_i^{*(k)} = \sum_{j^x=1}^{|\mathbf{x}|} \mathbb{1}\Big(i_{j^x}^{*(k)} == i\Big) + \sum_{j^e=1}^{|\tilde{\mathbf{x}}|+N_{\text{not-obs}}^{*(k)}} \mathbb{1}\Big(1 + i_{j^e}^{*(k)} == i\Big), \tag{S53}$$

where $i = 0, 1, \ldots$ and $\mathbb{1}$ is the indicator function that evaluates to 1 if $i_{j^x}^{*(k)}$ is equal to $i$, and otherwise to zero. The median estimate and 95% credible intervals of $x_i^*$ are obtained by drawing posterior samples $(k)$, repeating the calculation of (S53) for each $k$, and then summarising the set of samples.

Figure S24 shows the observed growth of subgraphs in red next to the predicted actual growth of subgraphs in blue (with 95% credible intervals) for Amsterdam MSM and heterosexuals.

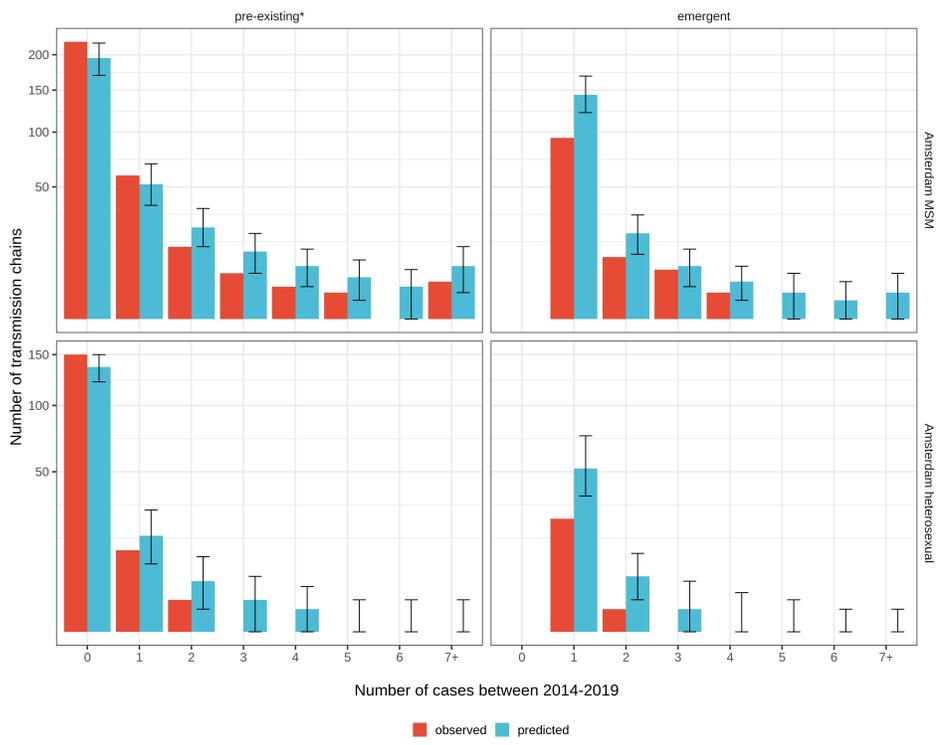

Figure S24: Growth of pre-existing (left) and emergent (right) phylogenetically observed subgraph sizes using estimated date of infection. * pre-existing prior to 2014.

## S7.2 Estimated origins of transmission chains between 2014-2018

If $\hat{\pi}_r = (\hat{\pi}_1, \ldots, \hat{\pi}_R)$ are the proportion of phylogenetically observed subgraphs since 2014 with geographic origin $r$, we can predict the origins of the pre-existing and emergent transmission chains, $y_j$, for each Monte Carlo sample through,

$$y_j^{(k)} \sim \text{Categorical}(\hat{\pi}_r). \tag{S54}$$

We then denote the proportion of predicted emergent transmission chains since 2014 with Amsterdam origin ($A$) by

$$\lambda^{(k)} = \sum_{j^e=1}^{|\tilde{\mathbf{x}}|+N_{\text{not-obs}}^{*(k)}} \frac{\mathbb{1}\left(y_{j^e}^{(k)} == A\right)}{\sum_{r=1}^{R} y_{j^e r}^{(k)}}. \tag{S55}$$

Table S5 reports the estimated ancestral origins of viral lineages in Amsterdam in the central phylogenetic analysis, and uncertainty estimates obtained from the bootstrapped analyses. The estimated origins predicted from the model are also reported, with 95% credible intervals.

## S7.3 Estimated number of new cases between 2014-2018

From Section S7.1, we can use the posterior predictive distribution of the number of new cases for a chain of a given index size (S46) to obtain a Monte Carlo prediction of the number of city-level cases since 2014. This is given by

$$x^{*(k)} = \sum_{j^x=1}^{|\mathbf{x}|} i_{j^x}^{*(k)} + \sum_{j^e=1}^{|\tilde{\mathbf{x}}|+N_{\text{not-obs}}^{*(k)}} (1 + i_{j^e}^{*(k)}). \tag{S56}$$

## S7.4 Estimated ethnicity of new cases between 2014-2018

For Amsterdam MSM, we consider geographic regions of birthplace of new cases $r \in$ {Netherlands; W. Europe, N. America & Oceania; E. C. Europe; S. America & Caribbean;

Other}. For Amsterdam heterosexuals, we consider georegions $r \in$ {Netherlands; S. America & Caribbean; Sub-Saharan Africa; Other}. Consider $\hat{\theta}_r = (\hat{\theta}_1, \ldots, \hat{\theta}_R)$ are the proportion of diagnosed individuals estimated to have been infected since 2014, born in geographic region $r$. We then predict the birthplace region $r$ of the total $x^{*(k)}$ new cases between 2014-2018, $z_n$, for each Monte Carlo sample through,

$$z_n^{(k)} \sim \text{Multinomial}(\hat{\theta}_r, x^{*(k)}). \tag{S57}$$

## S7.5 Proportion of locally acquired infections

The proportion of locally acquired infections since 2014 is defined by the proportion of city-level cases since 2014 that acquired infections in Amsterdam. In the model, all secondary infections originating from index cases of pre-existing transmission chains are infections that were acquired locally. Similarly, all secondary infections originating from index cases of emergent transmission chains are infections that were acquired locally. The index cases of pre-existing transmission chains do not contribute to the denominator (S56) because they existed prior to 2014. This leaves the index cases of emergent transmission chains, for which we also have a Monte Carlo estimate,

$$|\tilde{\mathbf{x}}| + N_{\text{not-obs}}^{*(k)}. \tag{S58}$$

A proportion of these index cases also acquired infection locally, from other risk groups in Amsterdam. We denote this proportion by $\lambda$ ((S55)). This allows us to estimate the proportion of locally acquired infections since 2014 through

$$\gamma_{\text{local}}^{(k)} = 1 - \frac{\left(1 - \lambda^{(k)}\right)\left(|\tilde{\mathbf{x}}| + N_{\text{not-obs}}^{*(k)}\right)}{x^{*(k)}}. \tag{S59}$$

The median estimate and 95% credible intervals of $\gamma_{\text{local}}$ are obtained by drawing posterior samples ($k$), repeating the calculation of (S59) for each $k$, and then summarising the set of samples.

Table S7 presents the estimated proportion of locally acquired infections by subtype, and the quantities used to calculate from equation (S59).

| Risk group | Subtype | Chains of non-Amsterdam origin $(1-\lambda)$ | Phylogenetically observed emergent subgraphs $(|\tilde{\mathbf{x}}|)$ | Emergent transmission chains (unobserved) $(N^*_{\text{not-obs}})$ | Total emergent chains (partially observed + unobserved) $|\tilde{\mathbf{x}}| + N^*_{\text{not-obs}}$ | Individuals in pre-existing and emergent chains $(x^*)$ | Proportion of infections that are importations $\left(\frac{(1-\lambda)(|\tilde{\mathbf{x}}|+N^*_{\text{not-obs}})}{x^*}\right)$ | External importations $\left(\frac{(1-\lambda)(|\tilde{\mathbf{x}}|+N^*_{\text{not-obs}})}{x^*}\right)$ | Locally acquired infections $(\gamma_{\text{local}})$ |
|---|---|---|---|---|---|---|---|---|---|
| Amsterdam heterosexual | B | 78.6% [70.3-86.5%] | 12 [12-12] | 15 [5-32] | 27 [17-44] | 60 [36-98] | 0.47 [0.27-0.69] | 36.6% [21.1-54.3%] | 63.4% [45.7-78.9%] |
| Amsterdam heterosexual | Non-B | 93.1% [88.3-97%] | 14 [14-14] | 18 [7-37] | 32 [21-51] | 61 [38-98] | 0.54 [0.33-0.78] | 50.3% [31.1-71.9%] | 49.7% [28.1-68.9%] |
| Amsterdam MSM | B | 99.5% [98.6-100%] | 85 [85-85] | 53 [35-74] | 138 [120-159] | 442 [352-552] | 0.31 [0.25-0.39] | 31.1% [24.4-38.7%] | 68.9% [61.3-75.6%] |
| Amsterdam MSM | Non-B | 98.5% [94.4-100%] | 29 [29-29] | 15 [6-26] | 44 [35-55] | 114 [75-184] | 0.39 [0.23-0.58] | 37.9% [22.7-56.5%] | 62.1% [43.5-77.3%] |

Table S7: Input quantities used to estimate proportion of infections acquired locally in Amsterdam

We then seek to estimate the proportion of locally acquired infections by ethnicity. Until now, all generated quantities are calculated for each subtype, without specific indexing. To obtain estimates of locally acquired infections by ethnicity, we apply weights to the subtype-specific estimates of locally acquired infections, using the proportion of predicted individuals from each georegion with subtype B or non-B infections. If $\gamma^{(k)}_{\text{local},s}$ is the proportion of locally acquired infections for subtype $s \in \{B,\text{non-B}\}$, and we introduce indexing for the birthplace by subtype, we then calculate the proportion of locally acquired infections by place of birth $r$ as follows:

$$\gamma^{r(k)}_{\text{local}} = \sum_S \gamma^{(k)}_{\text{local},s} * \frac{\mathbb{1}\left(z^{(k)}_{sn} == r\right)}{\sum_S \mathbb{1}\left(z^{(k)}_{sn} == r\right)}. \tag{S60}$$

As before, the median estimate and 95% credible intervals of $\gamma_{\text{local}}$ are obtained by drawing posterior samples ($k$), repeating the calculation of (S60) for each $k$, and then summarising the set of samples.

Table S8 presents the characteristics of the observed phylogenetically observed subgraphs alongside the model estimates for the parameters of the branching process model, and the proportion of infections estimated to have been acquired through city-level transmissions by transmission group and subtype.

Figure S25 presents the estimated $\gamma_{\text{local}}^{r(k)}$ from Equation (S60) and the composition of subtypes among the predicted total new cases used to estimate locally acquired infections by ethnicity from the subtype-specific estimates.

| Risk group | Subtype | Phylogenetically observed | | | | Model estimates | | | |
|---|---|---|---|---|---|---|---|---|---|
| | | New cases | Subgraphs | Average new cases | Transmission chains | Average new cases | Effective reproduction number | Variance-to-mean ratio | Infections acquired in Amsterdam |
| Amsterdam MSM | B | 241 | 368 | 0.65 | 421 [403-442] | 1.06 | 0.27 [0.23-0.32] | 1.71 [1.27-2.4] | 68.9% [61.3-75.6%] |
| Amsterdam MSM | Non-B | 65 | 55 | 1.18 | 70 [61-81] | 1.68 | 0.41 [0.29-0.55] | 1.35 [1.01-2.61] | 62.1% [43.5-77.3%] |
| Amsterdam heterosexual | Non-B | 21 | 105 | 0.2 | 123 [112-142] | 0.51 | 0.17 [0.09-0.28] | 1.25 [1.01-2.98] | 49.7% [28.1-68.9%] |
| Amsterdam heterosexual | B | 23 | 86 | 0.27 | 101 [91-118] | 0.6 | 0.19 [0.12-0.31] | 1.25 [1.01-2.67] | 63.4% [45.7-78.9%] |

Table S8: Empirical results from partially observed subgraphs in phylogenetic trees, and model estimates based on complete transmission chains, adjusting for sampling (those in study with a sequence available) for new infections since 2014. Estimated reproduction number and proportion of locally acquired infections are also presented.

## S7.6  Assessing model fit

To assess model fit, we perform posterior predictive checks against the phylogenetically observed growth distribution of subgraphs for each transmission group and subtype. To keep notations simple, we drop in what follows the suffixes that indicate dependence on transmission group and subtype.

We previously described the phylogenetically observed growth distribution through the number of pre-existing subgraphs with $m$ index cases by 2014 and $i$ new cases since 2014, $x_{mi}$, and the number of emergent subgraphs since 2014 with $n$ new cases, $\tilde{x}_n$. To generate posterior predictions $x_{mi}^*$ and $\tilde{x}_n^*$, we index the pre-existing, phylogenetically observed subgraphs by $j^x = 1, \ldots, |\mathbf{x}|$. With regard to emergent transmission chains, for the purpose of assessing model fit, we index only the corresponding phylogenetically observed subgraphs, $j^e = 1, \ldots, |\tilde{\mathbf{x}}|$. In analogy to Equation (S46), we use the sampling-adjusted size equa-

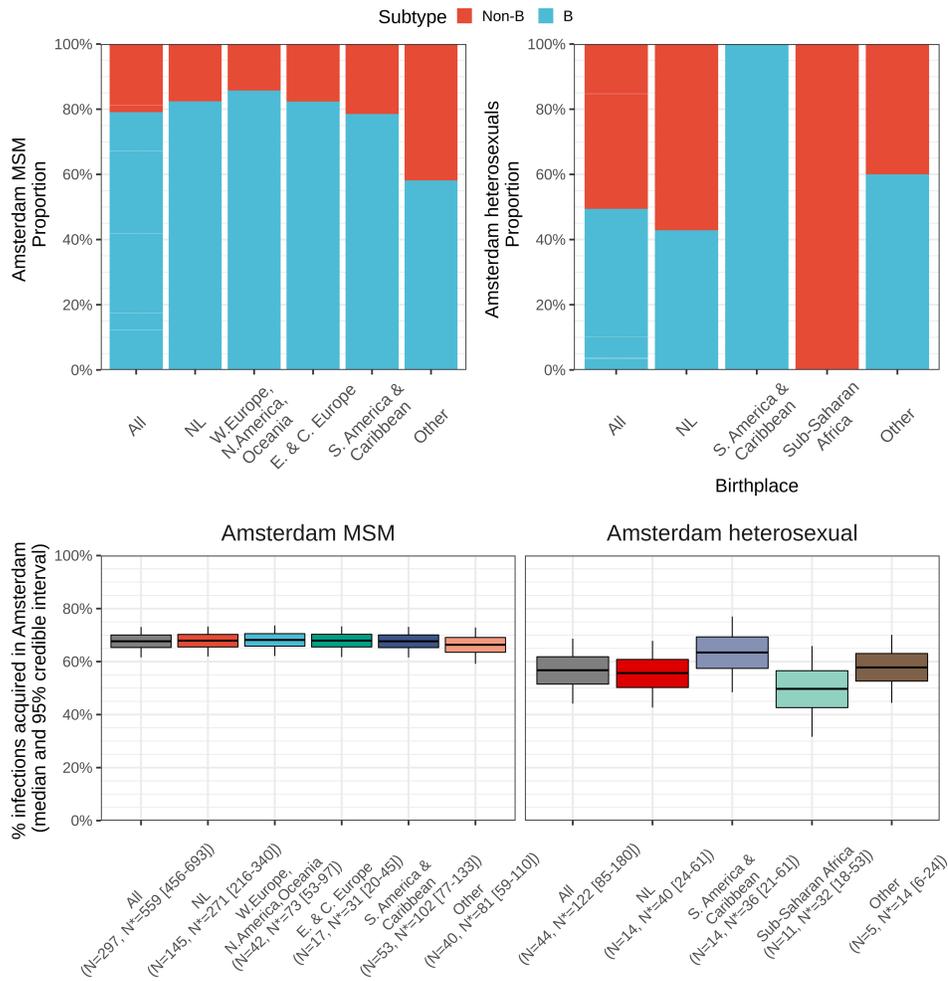

Figure S25: Top: Composition of subtypes among total predicted new cases. Bottom: Estimated local infections among MSM (left) and heterosexuals (right), stratified by place of birth between 2014-2018. N= number of sequences available, N*= estimated actual infections [95% credible interval].

tions (S33) and (S34), which lead to the posterior predictive densities

$$p_{\text{obs}}(i^*|\mathbf{x}, \tilde{\mathbf{x}}, m) = \int c_{\text{obs}}(i^*|m, \mu, \phi, \rho) p(\mu, \phi, \rho|\mathbf{x}, \tilde{\mathbf{x}}) d(\mu, \phi, \rho) \tag{S61}$$

$$p_{\text{obs}}(n^*|\mathbf{x}, \tilde{\mathbf{x}}) = \int \tilde{c}_{\text{obs}}(n^*|m=1, \mu, \phi, \rho) p(\mu, \phi, \rho|\mathbf{x}, \tilde{\mathbf{x}}) d(\mu, \phi, \rho). \tag{S62}$$

We use these posterior predictive densities to predict the (observed) growth of the pre-existing subgraphs, and the (observed) growth of the emergent subgraphs, and compare these predictions to the observed values. Specifically, given a Monte Carlo sample $\mu^{(k)}, \phi^{(k)}, \rho^{(k)}$ from the posterior distribution, we predict the growth of the pre-existing, phylogenetically observed subgraph $j^x$ through

$$i^{*(k)}_{j^x} \sim c_{\text{obs}}(\cdot|m_{j^x}, \mu^{(k)}, \phi^{(k)}, \rho^{(k)}). \tag{S63}$$

Similarly, we predict the growth of the emergent, phylogenetically observed subgraph $j^e$ through

$$n^{*(k)}_{j^e} \sim \tilde{c}_{\text{obs}}(\cdot|1, \mu^{(k)}, \phi^{(k)}, \rho^{(k)}). \tag{S64}$$

Finally, we aggregate ((S63)-(S64)) to obtain a posterior prediction of the observed growth distributions,

$$x^{*(k)}_{mi} = \sum_{j^x=1}^{|\mathbf{x}|} \mathbb{1}\Big(i^{*(k)}_{j^x} == i \text{ and } m_{j^x} == m\Big) \tag{S65}$$

$$\tilde{x}^{*(k)}_n = \sum_{j^e=1}^{|\tilde{\mathbf{x}}|} \mathbb{1}\Big(n^{*(k)}_{j^e} == n\Big). \tag{S66}$$

The posterior predictive check then tests if the observed $x_{mi}$, $\tilde{x}_n$ lie within the 95% range of the posterior predictive samples $\{x^{*(k)}_{mi}, k=1,\ldots,K\}$ and $\{\tilde{x}^{*(k)}_n, k=1,\ldots,K\}$.

Figure S16 shows the posterior predictive check for the MSM and heterosexual models, respectively, by subtype. 100% of the observed subgraph counts fall within the 95% credible intervals of the predicted subgraph size distribution, indicating very good model fit.

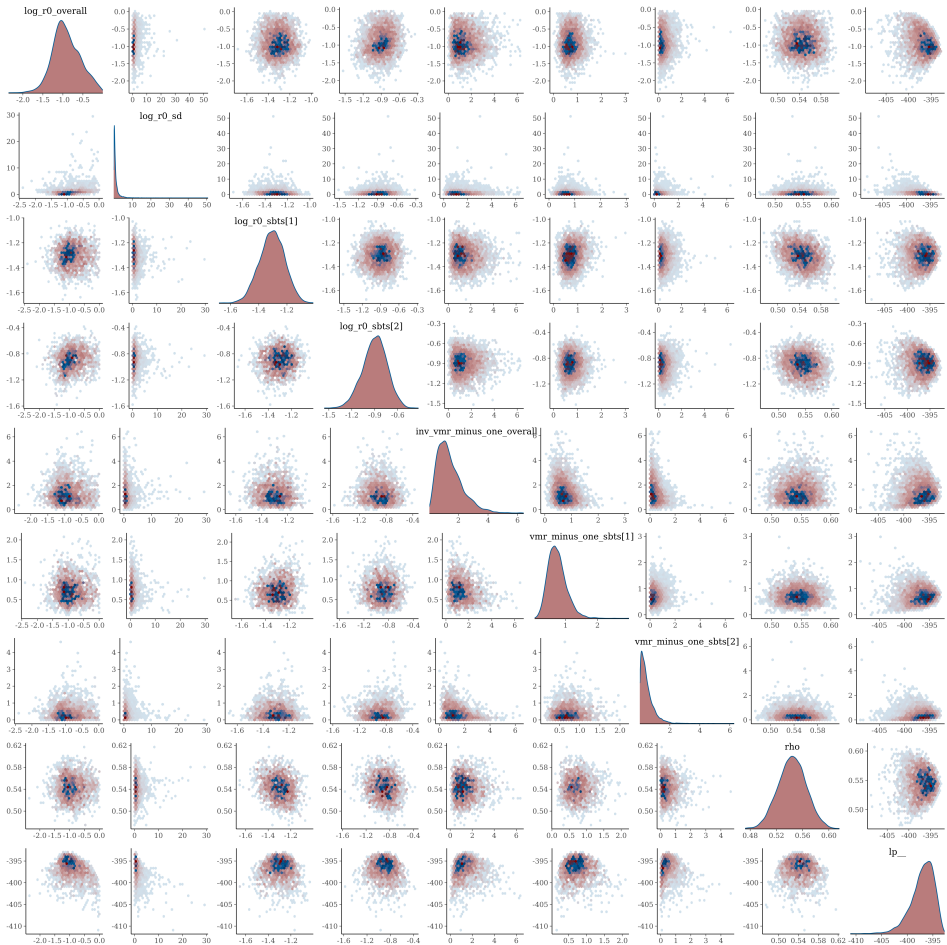

Figure S26: Pairs plot of the joint posterior density of the model parameters for Amsterdam MSM.

## S8 Sensitivity analysis

### S8.1 Observed subgraph size distribution considering infection date

Figure S28 shows how the observed growth distributions of the subgraphs compare when considering all diagnoses with a sequence between 2014-2018, and all diagnoses with a sequence estimated to have been infected between 2014-2018. There are fewer sequences in total when considering infection date, since some diagnoses since 2014 are estimated to be infections acquired before 2014.

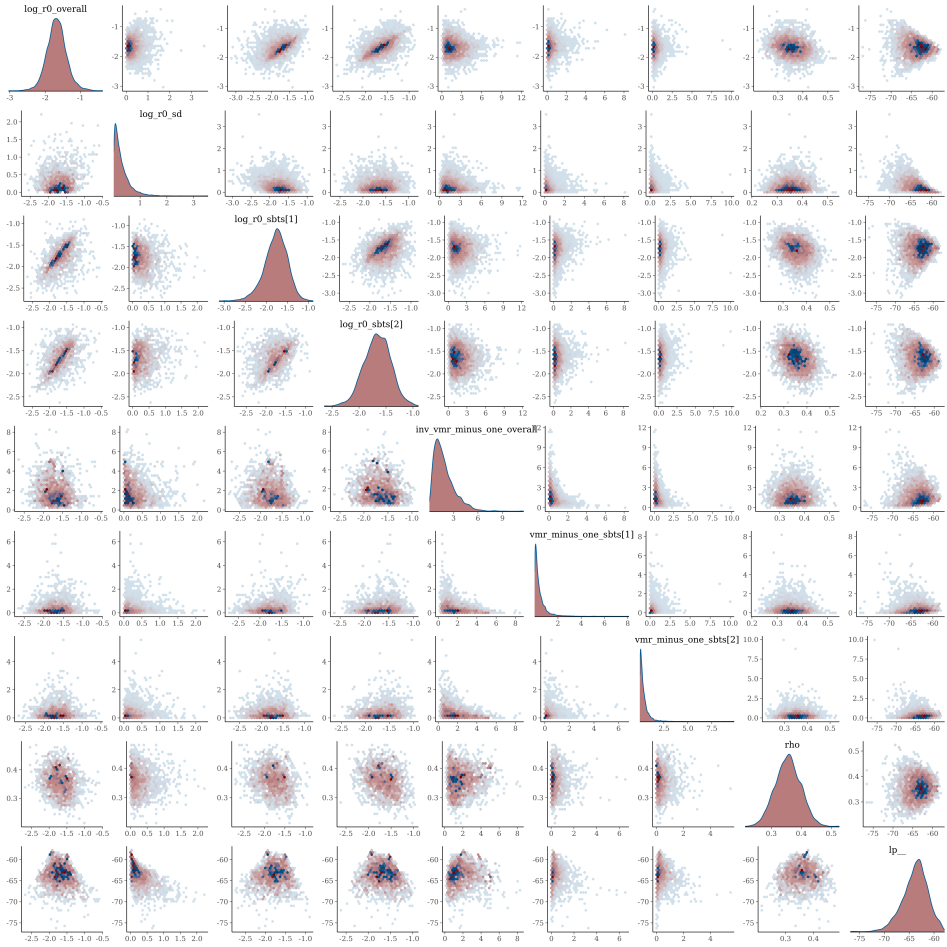

Figure S27: Pairs plot of the joint posterior density of the model parameters for Amsterdam heterosexuals.

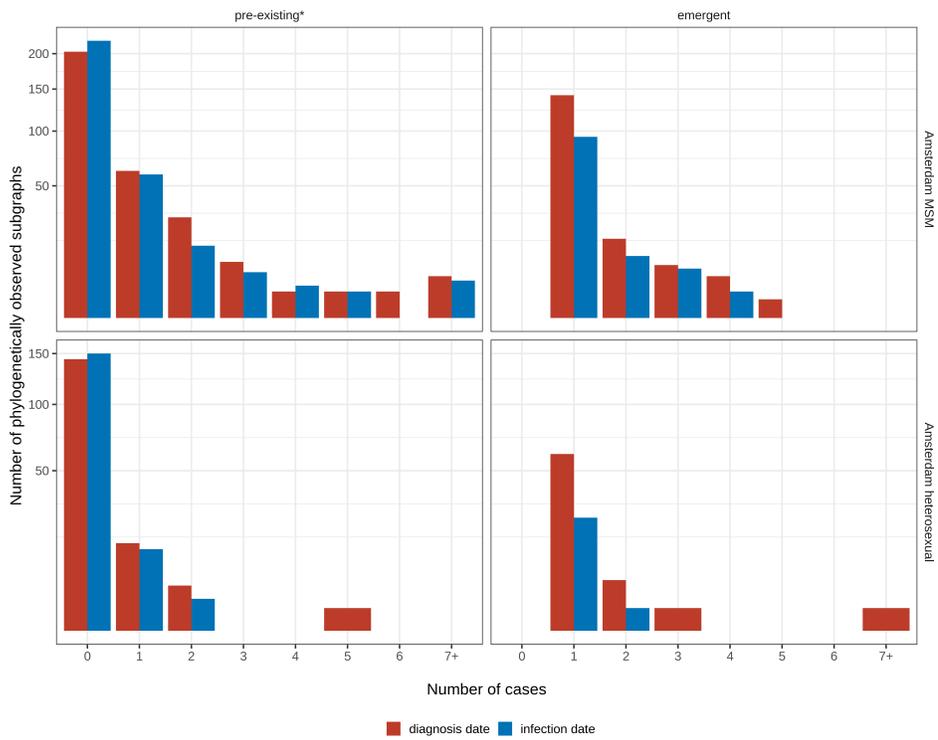

Figure S28: Growth of pre-existing and emergent phylogenetically observed subgraph sizes using diagnosis date and estimated date of infection. * pre-existing prior to 2014.